\documentclass[conference]{IEEEtran}
\IEEEoverridecommandlockouts
\usepackage{cite}
\usepackage{url}
\usepackage{hyperref}
\usepackage{pgfplots}
\usepackage{tikz}

\hypersetup{
    colorlinks=true,        
    linkcolor=black,        
    citecolor=blue,         
    filecolor=magenta,      
    urlcolor=cyan           
}
\usepackage{amsmath,amssymb,amsfonts}
\usepackage{algorithmic}
\usepackage{graphicx}
\usepackage{textcomp}
\usepackage{xcolor}
\usepackage{tabularx}
\def\BibTeX{{\rm B\kern-.05em{\sc i\kern-.025em b}\kern-.08em
    T\kern-.1667em\lower.7ex\hbox{E}\kern-.125emX}}
\begin{document}
\title{A Passwordless MFA Utlizing Biometrics, Proximity and Contactless Communication
}

\author{\IEEEauthorblockN{Sneha Shukla, Gaurav Varshney, Shreya Singh, Swati Goel}
\IEEEauthorblockA{\textit{Department of CSE, IIT Jammu} \\
sshukla19.01.02@gmail.com, \{gaurav.varshney, 2022pct0019, iitjmu81166\}@iitjammu.ac.in}


}

\maketitle

\begin{abstract}
Despite being more secure and strongly promoted, two-factor (2FA) or multi-factor (MFA) schemes either fail to protect against recent phishing threats such as real-time MITM, controls/relay MITM, malicious browser extension-based phishing attacks, and/or need the users to purchase and carry other hardware for additional account protection. Leveraging the unprecedented popularity of NFC and BLE-enabled smartphones, we explore a new horizon for designing an MFA scheme. This paper introduces an advanced authentication method for user verification that utilizes the user’s real-time facial biometric identity, which serves as an inherent factor, together with BLE- NFC-enabled mobile devices, which operate as an ownership factor. We have implemented a prototype authentication system on a BLE-NFC-enabled Android device and initial threat modeling suggests that it is safe against known phishing attacks. The scheme has been compared with other popular schemes using the Bonneau et al. assessment framework in terms of usability, deploy ability, and security. 
\end{abstract}


\section{INTRODUCTION}
Phishing attacks continue to play a dominant role in the digital threat landscape, as mentioned by the APWG (Anti-Phishing Working Group), 2020 \cite{report3}. In traditional phishing attacks, attackers lure victims and make them input their credentials on a look-alike sign-in page of a popular site that they have targeted. However, this technique became less efficient with the adoption of the two-factor authentication (2FA) \cite{report29}\cite{report30} scheme, which requires two authentication factors for successful authentication. There are now many new ways to get around 2FA, some of which use MITM (Man-in-the-Middle) phishing attacks like RT (Real Time) and CR (Control Relay) MITM phishing attacks \cite{rt1} and \cite{rt2}. While 2FA provides another layer of security, it primarily protects against static or long-term credentials (mostly usernames and passwords) stealing attacks. Adding a second dynamic factor, such as an OTP, thwarts any attacks where the attacker has access to a static user credential. It does not protect against other types of phishing attacks, like MBE (malicious browser extension) \cite{report31}\cite{report32}. These malware extensions often possess legitimate functionality, yet they serve as a conduit for data theft, including the capture of any second factor used for login. If the user enters all credentials, whether dynamic or static, for identity verification, and attackers have a proven trivial way to capture them in real-time, any authentication scheme is susceptible to compromise.
The above conditions are likely to be satisfied by existing, highly sought-after web-based authentication techniques. Hence, the assailant has the potential to pilfer the user's identity data and gain unauthorized access to their confidential information. This paper introduces a cutting-edge and advanced authentication method for user verification. It utilizes the user's real-time facial biometric identity, which serves as an inherent factor, together with BLE-NFC-enabled mobile devices, which operate as an ownership factor. Before final authentication, it is necessary to pair two devices that are equipped with Bluetooth Low Energy (BLE) capability to establish an association between them. Using linked devices provides a reliable method for effectively thwarting web-based phishing assaults on authentication methods in real-time.
Throughout this paper, the term "first device" refers to the mobile device attempting to authenticate some system and is a registered device. The term "second device" refers to the device on which login is being performed using the first device. Also, the “challenge” term represents a required action to complete authentication. The first authentication step employs the MobileFaceNet model\cite{report15}, which is an extremely efficient Convolutional Neural Network (CNN) model tailored for high-accuracy real-time face verification on mobile and embedded devices, utilizing less than 1 million parameters. Trained on datasets like MS-Celeb-1M with the ArcFace loss function\cite{Deng_2022}, MobileFaceNets achieve superior accuracy rates, comparable to state-of-the-art big CNN models, while offering significant speed improvements with actual inference times as low as 18 milliseconds on a mobile phone. Its carefully designed architecture overcomes the weaknesses of common mobile networks for face verification, making it a groundbreaking solution for applications requiring efficient and accurate face recognition capabilities on mobile platforms. In this face images are preprocessed using MTCNN to detect landmarks and align faces, then resized to a standard size (e.g., 112x112 pixels). These preprocessed images are passed through a MobileNetV2 CNN to extract high-level features, using a global depthwise convolution layer (GDConv) instead of global average pooling to capture more discriminative features. The output of the CNN is a feature vector representing the face. For face verification, feature vectors from two images are compared using similarity metrics here Euclidean distance is used because it is proven to give good results for face-matching problems. If the similarity exceeds a threshold, the faces are considered the same, otherwise, they are different. This process allows MobileFaceNets to perform accurate face verification, leveraging GDConv for enhanced feature discrimination. As a challenge, a Nonce and Bluetooth address are used. When the first device tries to log in to the second device, the first device captures the user's real-time face biometric using the MobileFaceNet  model and generates a real-time dynamic face-bio URL for the login phase. Then the user needs to perform an NFC tap from the first device against the second device to transfer the token, which is then verified by the server, which accesses the same token from the second device before verification. Finally, the second device checks whether the first device is in its proximity to authenticate the identity of the first device. To test the working prototype, two BLE-NFC-enabled smartphones and an Android application were developed. We outline the paper's main contribution below:
\begin{enumerate}
    \item The security of existing multi-factor authentication schemes has been analyzed against the latest attacks that an attacker carries out via RT/CR MITM and MBE-based phishing. Our analysis indicates that most of the popular schemes are not capable of handling these attacks. Only a handful of them, such as Yubikey U2F, can mitigate these attacks, but they require their user to purchase a hardware key or token and always carry this additional hardware with them.
    \item We propose a secure password-less authentication scheme, that can handle RT MITM, CR MITM, and MBE-based phishing attacks and uses a BLE-NFC-enabled smartphone, a device generally owned by every user using the Internet.
    \item We verify the proposed authentication protocol using an automated security protocol verification tool.
\end{enumerate}

The remainder of the article is organized as follows:
In Section 1, the phishing attacks that are popular on the web are explained, along with the reason for study in this area. It also lists the important points of the paper. In Section 2, we talk about the well-known and current multi-factor authentication methods that are used for web authentication. We also focus on the proposals that are said to be able to defend against RT MITM, CR MITM, and MBE-based phishing attacks. This part also lists the study gaps that have been found. In Section 3, we talk about the threat model, assumptions, and abbreviations used, as well as the design and how the suggested secure password-less authentication scheme works. This fills in the research gaps that were found in Section 2. Section 4 provides details of the implementation and performance evaluation. Section 5 discusses the security evaluation of the proposed scheme. Section 6 highlights the modeling and verification of the authentication protocol and shows the experimental results. It also shows a comparison of authentication systems in terms of usability and shows comparison with the existing multi-factor schemes in terms of usability, deployability, and security using the Bonneau et al. framework. We verified the authentication scheme using an automated security protocol verification tool. As an initial step, the protocol was modeled using a high-level protocol specification language to verify and check whether secrecy and authenticity properties were violated or not. Section 7 discusses the proposed model's key limitations and concludes with discussions on future work possible in the area.

\section{RELATED WORK}
In this section, we particularly review 2FA mechanisms, face recognition, and Bluetooth based - proximity authentication schemes, that are proposed to handle the recent phishing attacks.

\textbf{Face recognition using mobile}: In 2015, Xie et al. \cite{camauth} proposed CamAuth, a 2FA scheme that leverages the user’s mobile as a second authentication factor. The approach claims to solve the problem of MITM phishing attacks and utilizes a combination of Diffie-Hellman keys exchanged between the client browser and the server to prove the user's identity, and then verify it using the user’s mobile device via using both the user's PC and mobile cameras to exchange data that is encapsulated within a QR code. Phoolproof \cite{phool} is a public-key-based scheme for strengthening the bank transaction system. The user is required to choose a bank site from the whitelist on the phone and then wait for information exchange between the phone and the PC. The approach claims to thwart Man-in-the-Middle attacks after setup and protects a user’s account even in the presence of keyloggers. In 2018, Zavrik et al. [1] proposed a 2FA web authentication scheme based on facial recognition to derive OTP and use this OTP to generate mobile device number IDs for future login sessions. Recently, Xie et al. proposed CamTalk, a light-based communication framework for bidirectional secure information transfer between smartphones by leveraging the smartphone's display-camera channel \cite{camtalk}.

\textbf{Short-range communications}: Bluetooth, WiFi, or NFC are also widely adopted to support two-factor authentication. In 2019 \cite{ali}, Ali et al. proposed web authentication using Bluetooth devices that authenticate users by checking the presence of known or expected Bluetooth devices in the proximity of the user, which can either be explicitly specified by the user or implicitly learned by the system through previous successful logins. Saxena et al. proposed a short-range device pairing protocol, VIC (Visual authentication based on Integrity Checking), which is also based on a unidirectional visual channel \cite{saxena}. The pairing process requires the use of another wireless communication channel, such as Bluetooth. ViC is suitable for web authentication. An authentication service provider, SAASPASS \cite{report19}, leverages location-based iBeacon Bluetooth Low Energy (BLE) technology to authenticate users via Bluetooth communications between their registered phones and nearby login computers. Similarly, another 2FA proposal, PhoneAuth \cite{phoneauth}, sets up unpaired Bluetooth communications between a login computer and the user's phone via Bluetooth using a new challenge-response protocol.
As NFC is widely embedded into today’s commodity smartphones, Facebook \cite{report34} introduced a physical NFC security key that allows users to log in to their accounts on their smartphones via NFC. This solution makes the hardware token-based two-factor authentication process faster. Instead of reading an authentication code from a hardware token and inputting it to a login computer, a user just taps an NFC security key against his or her smartphone so as to complete an authentication session. However, this solution requires additional hardware and its cost is of similar concern as in the case of hardware-token-based 2FA.

\textbf{Hardware Token}: Hardware-token-based 2FA is a widely deployed 2FA solution in practice (e.g., in the financial industry). It requires users to carry and use hardware tokens for authentication. Other hardware security tokens, such as RSA SecurID, are used during an authentication session to generate an authentication code at fixed time intervals (usually 60 seconds) according to a built-in clock and a factory-encoded random key (known as "seed"). After entering the first factor, a user reads the authentication code from the hardware token and inputs it to a login computer. Hardware-token-based 2FA requires users to interact with their hardware tokens. It also requires a service provider to manufacture a number of hardware tokens and distribute them to all customers. In double-armored Tricipher \cite{report35}, multipart credentials are used. In this scheme, one part of the credentials is stored in a secure enterprise data center, and the other is with the user. Also, the security key is stored on the user’s device and is known to the server. Both the username and password entered by the user are encrypted using this key only. The enterprise data center encrypts (signs) user input with the part of the credential available on it and sends this encrypted message to the user. The server receives this encrypted message directly, completing the login process. The scheme needs an additional hardware security key during the login process. During web login, Yubikey uses the U2F protocol to authenticate the user. Several companies, including Lenovo and PayPal, formed the Fast Identity Online (FIDO) Alliance in 2012. The FIDO alliance aims to bring different authentication schemes together by providing a set of standards that simplify their adoption and use in web authentication. The FIDO website \cite{report36} has recently published its specifications.FIDO CTAP, which stands for Fast Identity Online Client to Authenticator Protocol, is a standard
developed by the FIDO (Fast Identity Online) Alliance. It’s designed to enhance online security by enabling passwordless authentication. Traditionally, online authentication often relies
on passwords, which can be vulnerable to various attacks such as phishing, brute force, and password theft. FIDO CTAP aims to address these vulnerabilities by introducing a more secure and convenient authentication method. Instead of passwords, FIDO CTAP \cite{wiki_ctap} utilizes cryptographic keys stored on dedicated hardware security keys or devices, such as USB tokens or smartphones. These keys are used to verify the user’s identity without transmitting sensitive information over the internet.



\begin{table*}
\caption{Abbreviations Used}
\label{tab:abbr}
\begin{tabularx}{\textwidth} { 
   >{\raggedright\arraybackslash}X 
   >{\raggedright\arraybackslash}X
   >{\raggedright\arraybackslash}X
   >{\raggedright\arraybackslash}X  }
 \hline
\textbf{Variable} & \textbf{Description} &\textbf{Function} & \textbf{Description} \\
\hline 
\( EM \) & Email Address & HTTPS(DATA) & HTTPS message carrying data. \\
\hline 
\( PWD \) & Password & GEN(x) & Generate a cryptographically secure random secret \( x \).\\
\hline 
\( FBIO \) & Real-time face biometric & \( CMP_{DB}(x,y,z....) \) & Matches user credentials received as arguments with those present in the server database.\\
\hline 
\( FbURL \) & Real time face biometric URL & STORE(x)  & Stores the value ‘x’, encrypted (symmetric encryption AES) with \( SK \) and SALT in Android App storage. \\
\hline 
\( AID \) & An ID stored in App that identifies App installation uniquely over a phone  & SAVE\textsubscript{DB} & Adds a new tuple (EM, PWD, AID, SALT, FbURL) to the server database. \\
\hline 
\( SK \) & Shared secret between device and server & REPLACE(a,b) & Replaces the encrypted value ‘a’ with encrypted value ‘b’ in server database storage. \\
\hline 
\( BT\textsubscript{i} \) & Bluetooth Address of device i(log in or registered device) & SEARCH(x) & Search the x value in the nearby proximity of the device using Bluetooth technology. \\
\hline 
\( K\textsubscript{s} \) & Secret to encrypt the challenges and decrypt the response & NFC(x) & Tap the encrypted token against the second device using NFC. \\
\hline 
\( N\textsubscript{1} \) & A random 10-digit long string generated for verification of registered device & W\textsubscript{DB}(x) & Creates a new column in the server database and stores value ‘x’. \\
\hline 
\(S\) & A random salt \\
\hline 
\( MATCH \) &  A predefined string that conveys a specific message. \\
\hline

\end{tabularx}
\end{table*}

\section{Proposed Scheme}
\subsection{Threat Model}
In our proposed solution we have taken into consideration an attacker having the following set of capabilities:
\begin{enumerate}
\item \textbf{Phishing, MITM Phishing}: The attacker can trick victims into entering their credentials on the phishing website and use those credentials to gain access to their account. Additionally, a phisher can deploy remote desktop relay/capturing modules (Ulterius, Teamviewer, etc.) on the victim machine or utilize QRLjacking \cite{report24} techniques to carry out RT MITM phishing and CR MITM phishing attacks.\item \textbf {MBE Attacks}: The attacker can carry out this attack by installing a malicious browser-based extension that provides legitimate functionality in the foreground and performs stealthy activities such as logging user keystrokes and sniffing user-entered information by seeking the same set of user permissions for both the foreground and background, respectively.
\item \textbf {App Spoofing}: The attacker can install a spoofed Android app on the user's first device and lure them into entering their account information over apps \cite{spoof}.
 \item The phisher is capable of the following :
    \begin{itemize}
        \item The attacker can sniff \( \mathnormal{BT_{1}} \) and install spoofed Android apps and extensions during the Android registration phase. The attacker can log users' keystrokes with the help of \( \mathnormal{MBE} \). It is assumed that the user will be using a legitimate Android app, and the web registration will be free from any attacks like many other schemes \cite{camauth}.\item The scheme assumes that data communication between the browser (client) and the web server over \( \mathnormal{HTTPS} \) is safe from sniffing, and it can be used as a secure channel to exchange secret keys. Also, web servers and databases are assumed to be safe from any form of attack.\item The attacks carried out by hijacking sessions, by malformed DNS, or by host-based malware are not within the work's current scope, and it is assumed that the organization provides the secure  \(\mathnormal{DNS} \) service. Also, the attacks launched due to a modified source code browser installed on the user's laptop are not in the current scope of the work.
    \end{itemize}

\end{enumerate}

\subsection{Assumptions}
\begin{itemize}
    \item Both mobiles and Desktop are assumed to have inbuilt BLE and NFC capabilities, which are more efficient than carrying an extra security key or token. The first device is NFC and BLE enabled, and it is in proximity to the second device.
    \item It is assumed that the user logs into the app installed on the second device using the app deployed on the first device.
    \item The user can use his first device (the one on which the app is installed) as a registered device during the web login phase.
   
\end{itemize}

\subsection{Abbreviations used}
Before presenting the scheme, we first
introduce important notations for clarification
purposes summarized in the Table~\ref{tab:abbr}.
\begin{figure*}
\centering
\includegraphics[width=16cm, height=9cm]{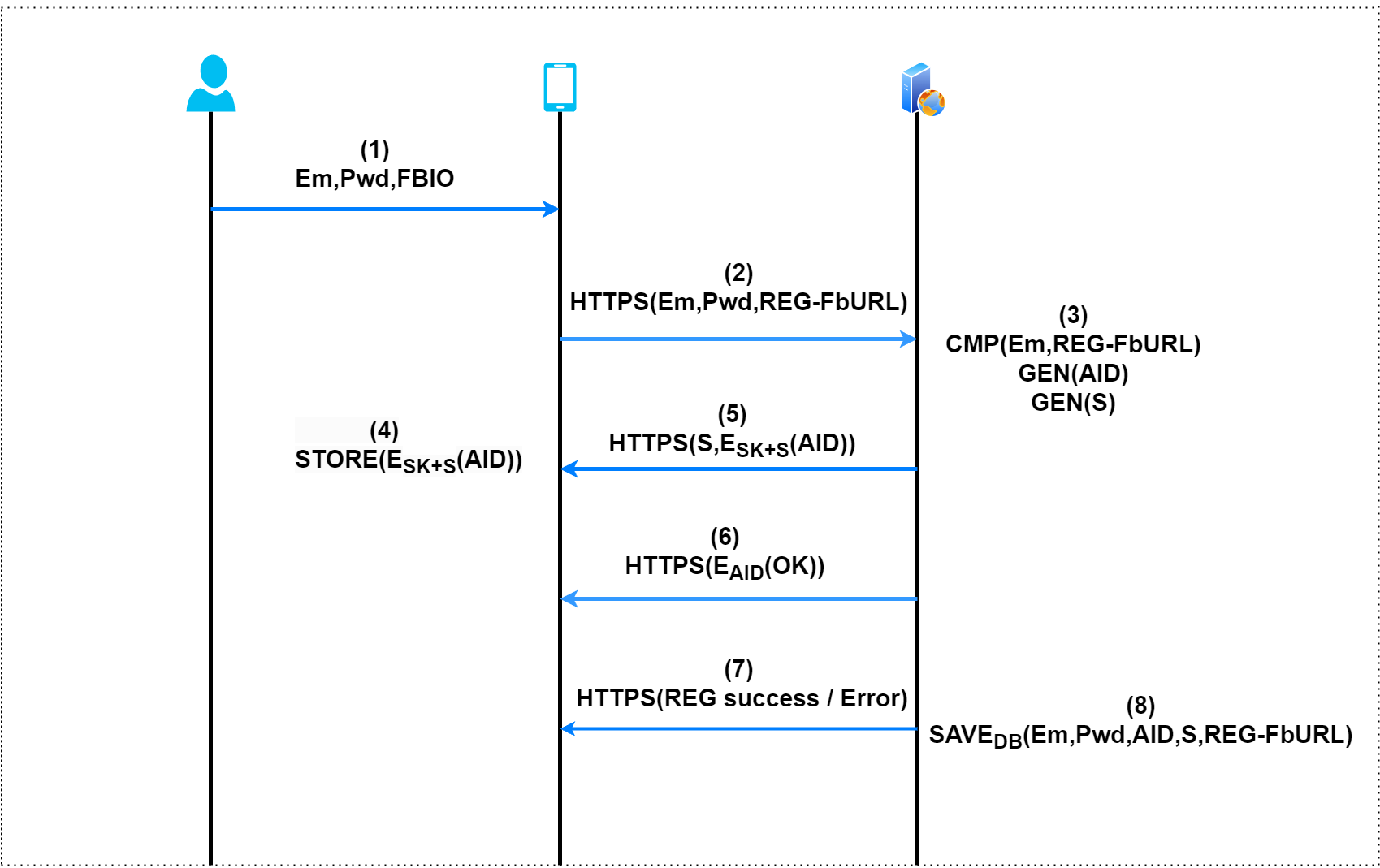}
\caption{Smartphone Registration Phase}
\label{fig:reg}
\end{figure*}

\subsection{System Overview}
The proposed authentication scheme addresses some of the research gaps identified in the previous section. A user may log in to a website through his Smartphone under the proposed scheme. A Smartphone and an Android app (or iPhone app) installed on it are required. The proposed multi-factor authentication scheme requires a real-time face biometric, a device identification token (Bluetooth address), and an instance ID of the Android app. Unlike other schemes where the user enters his login username manually on the system. The app generates a random secret number (N1) and retrieves the Bluetooth address (BT1) of the device, updating these details in the website's database linked to the user's email address. Next, the app communicates with the web server over HTTPS, providing an encrypted identifier \( \mathnormal{E_{N_{1}}(E_{SK + S}(AID))}\) and the email (EM) for device identification. The server verifies this communication by encrypting the stored identifier (AID) with a secret key (SK) and salt (S), then matching it with the received\( \mathnormal{E_{N_{1}}(E_{SK + S}(AID))}\). Upon successful verification, the server sends the salt (S) to the app. The app decrypts the AID using the received salt (S) and its stored SK, creating an encrypted authentication string containing user details (Email,Login-FbURL,SID). This string is shared via NFC tap to a desktop for further authentication within a limited timeframe. The desktop then sends this string and its Bluetooth address \(\mathnormal{BT_{2}}\) to the server for identity verification. The server decrypts the string using stored user data, verifying the biometric match between login and registration face biometric. If successful, the server sends a MATCH message encrypted by \(\mathnormal{BT_{2}}\) to the desktop, which then searches for \(\mathnormal{BT_{1}}\) in proximity. Upon finding \(\mathnormal{BT_{1}}\), the desktop encrypts a token with \(\mathnormal{BT_{1}}\) and sends it back to the server for AID' generation. The server encrypts AID' and sends it with salt (S) back to the app, which updates the authentication process. Finally, the app generates a final key \(\mathnormal{(E_{SK + S}(AID))}\) using the current AID and sends it for verification. After successful key verification, the server updates the database and sends a login success or error message. This complex yet secure process integrates face biometrics, Bluetooth identifiers, and encryption to ensure robust user authentication and account access.

\subsection{Protocol Details}
A user can log in to a website using the first device, which is a smartphone. The new user registers on the website during the registration phase for the first time using Email ID and Password. 
The proposed scheme includes two phases: (1) Android Registration Phase - First Device (ARP - FD) and (2) Android Login Phase - Second Device (ALP - SD)

\textbf{ARP - FD}: In ARP - FD  a new user registers on the website and logins the account over Smartphone.
The Smartphone App and the webserver have a shared secret key named SK. This secret key is shared over TLS/SSL prior to the smartphone registration phase.
\begin{enumerate}
   {\item The user enters an email address (Em) and password (Pwd) in the first step. After submitting, the camera opens up on the Smartphone App to take the real-time face biometry (FBIO) of the user. This biometric information is then transmitted to the server. The server processes and stores this information securely. During this storage process, the server generates a unique URL, referred to as FbURL, which serves as an address or link to the stored biometric information. This URL is created by the server to provide a specific and direct path to access the user's biometric data. 
   
   \item The user clicks the Sign Up button to submit the user details from the Smartphone App to the web server over HTTP.
   
   \item The web server checks if the email ID chosen by the user is unique and then generates two secret tokens, AID and Salt (S). The server then encrypts the AID using a key that is a combination of SK and salt (S). 
   \item The server then sends \( \mathnormal{S, E_{(E_{SK + S}(AID))}}\) to the Smartphone App in the final response.
   
   \item The Smartphone App stores the encrypted AID locally in its app storage. The app decrypts the AID using SK and SALT (S) to complete step 6.
   
   \item To confirm that the Smartphone App has received an AID, the App encrypts a stringified fixed token with the AID, which is already known by the server to signify approval, and then sends it to the web server.
   
   \item Once the web server receives the valid stringified fixed token encrypted via the correct AID, it sends a registration success or error message.
   \item Then it creates a new database entry for the user, stores Em, AID, Pwd, and FbURL, and sends a registration success message back to the app.
   After this, the app discards Salt (S) and AID from its storage and only stores the encrypted AID in its local storage for further authentication.
   After the registration phase step, the Smartphone App has encrypted AID in its local storage and the SK.
}
\end{enumerate}
\begin{figure*}
\centering
\includegraphics[width=19cm, height=16cm]{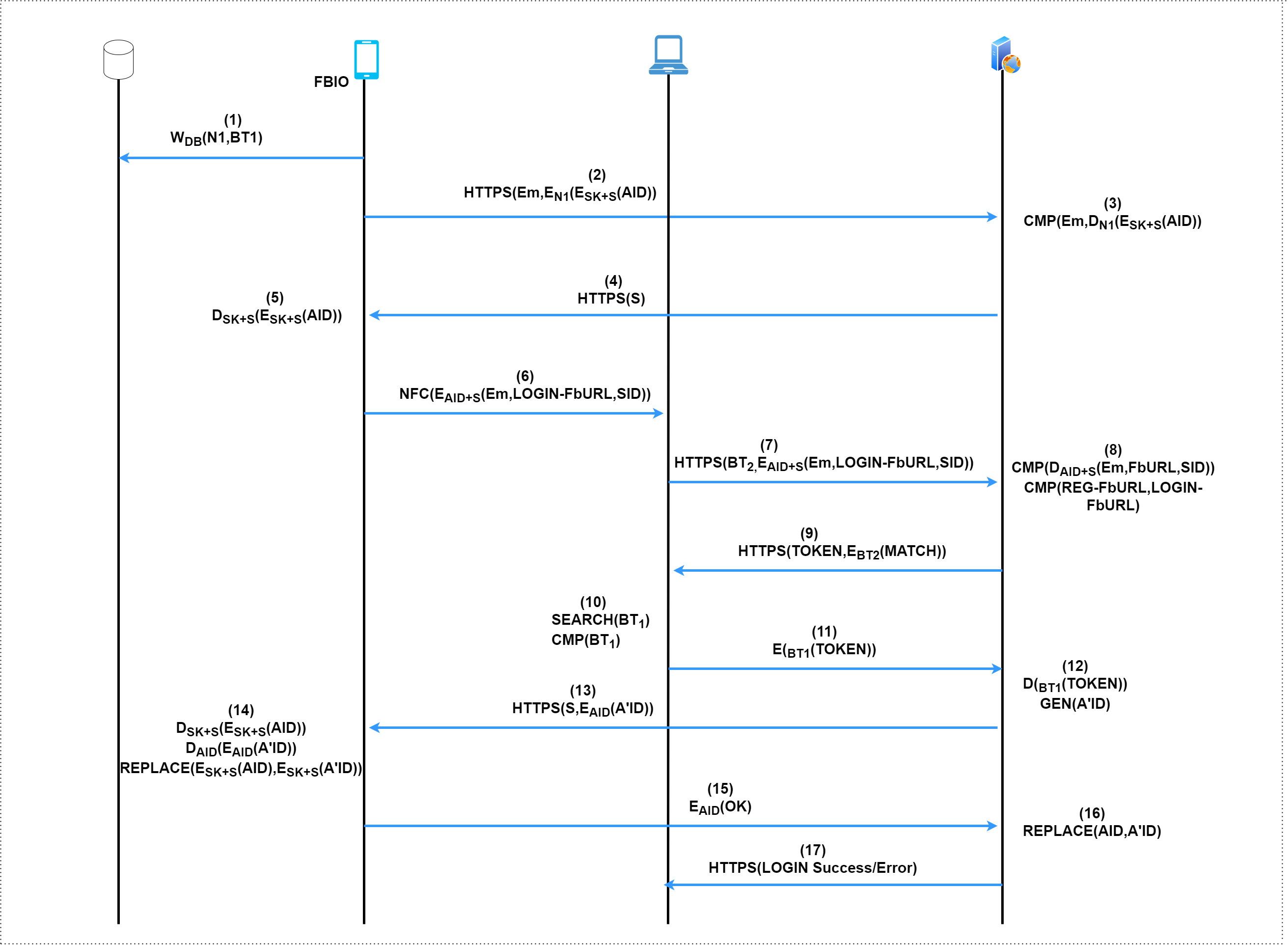}
\caption{Smartphone Login Phase}
\label{fig:login}
\end{figure*}

The messages exchanged during ARP - FD is shown in Fig ~\ref{fig:reg}. Our main contribution is ALP - SD which describes a method to authenticate a second device using the first device.

\textbf{ALP - SD}: To log in to his registered account on the second device, the user must be logged in to his Smartphone App account on the first device. Logging over the Smartphone App is needed for this authentication model to take real-time face biometrics and transfer challenges to the second device through an NFC tap to log the second device.

The user should now follow the below steps to log in to his registered account using the second device:
\begin{figure*}
    \centering
    \includegraphics[width=14cm, height=7cm]{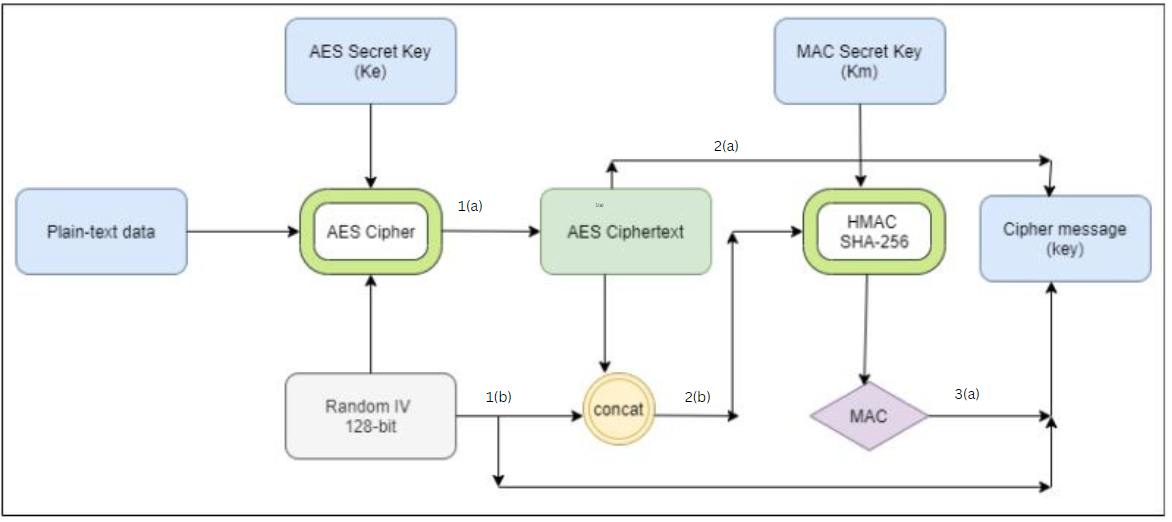}
    \caption{Encrypt-then-MAC-Sequential-scheme}
    \label{fig: encrypt mac}
\end{figure*}
\begin{enumerate}

\item During this phase, the user submits his or her biometrics by capturing his or her real-time face image using the built-in camera in the logged-in Smartphone App. The Smartphone App generates a random secret number \(N_{1}\) and retrieves the Bluetooth address \(BT_{1}\) of the smartphone device. It then updates (\(N_{1}\), \(BT_{1}\)) in the database of the website associated with the user's registered email (Em) address.

    \item After updating the database the Smartphone App initiates the communication with the web server and provides \( \mathnormal{E_{N_{1}}(E_{SK + S}(AID))}\) and email (Em) for the first device identification over \(HTTPS\).

  \item Web server verifies the user’s communication from the Smartphone App by encrypting the AID stored in the database with secret key (SK) and salt (S) and then encrypting it with nonce (\(N_{1}\)), to get the server copy of \( \mathnormal{E_{N_{1}}(E_{SK + S}(AID))}\).

  \item If received and the server copy matches, the web server sends  \(SALT\) over \(HTTPS\) to the smartphone app.

    \item The Smartphone App decrypts AID using the received salt (S) from the web server and its secret key SK which is already stored in the App. It then creates a concatenated authentication string in the form of \( \mathnormal{E_{AID+S}(Em, LOGIN-FbURL, SID))}\). 
    
    \item The user shares this concatenated authentication string using the Smartphone App via NFC tap to the second device which is a Desktop on which user authentication has to be achieved and starts the timer for the second device to complete the authentication before it expires. After successful sharing, the app on the first device deletes this concatenated authentication string from its local storage.

    \item The Desktop shares this concatenated authentication string and its Bluetooth address \(BT_{2}\) to the web server for its identity verification.

    \item The server will decrypt concatenated authentication string using the AID and salt present in its database for the user and obtain session-id (SID), Em, and LOGIN-FbURL and matches the biometric identity at the login time (LOGIN-FbURL) and registration time (REG-FbURL) if the match percent reaches a certain threshold the session id corresponding to the user is saved in the database. 
    
    \item If the biometric identity is matched successfully the message \( \mathnormal{MATCH}\) is sent to the Desktop which is encrypted by  \( \mathnormal{BT_{2}} \) along with random \( \mathnormal{TOKEN}\). \( \mathnormal{MATCH}\) is a predefined string that conveys a specific message. \( \mathnormal{TOKEN}\) is used for the protection of replay attacks.
    
    \item After receiving \( \mathnormal{MATCH}\) second device starts searching for \( \mathnormal{BT_{1}} \) in its proximity.
    
    \item After getting  \( \mathnormal{BT_{1}} \) it will encrypt the received \( \mathnormal{TOKEN}\)  using \( \mathnormal{BT_{1}} \) and send to web server. 
    \item The server decrypts this received message using \( \mathnormal{BT_{1}} \) if it is \( \mathnormal{TOKEN}\) then it will generate a new AID’.

    \item Web server encrypts AID’ using AID, then sends salt (S) and encrypted AID’ to the first device.
    
    \item Then first device extracts AID by decrypting locally stored encrypted AID using SK and salt (S). Then using AID first device decrypts the encrypted AID’ and using this AID’ it will replace \( \mathnormal{(E_{SK + S}(AID))}\) with  \( \mathnormal{(E_{SK + S}(AID'))}\).
    
    \item The Smartphone App then uses the AID of the current login session to generate  \( \mathnormal{(E_{AID}(OK))}\), and sends it to the server for key verification. The server decrypts the challenge with the old AID stored in its database and returns the user account over \( \mathnormal{HTTPS}\) connection.
    
    \item After key verification server replaces AID with AID’ in the database.
   \item Then the server sends a LOGIN success or error message.

\end{enumerate}
The messages exchanged during ARP - FD is shown in Fig ~\ref{fig:login}. The appendix shows a video made during our testing.

\subsection{AID, \( \mathnormal{N_{1}}\), Ks and SALT generation}
The AID and Ks are some of the authenticators that are generated using the flutter library, which is cross-platform string encryption that uses AES256 CBC +  PKCS5 + Random IVs + HMAC-SHA 256 to ensure confidentiality, integrity, and authentication of string. The \( \mathnormal{N_{1}}\) is generated using the \("random numeric"\) method, and \( \mathnormal{SALT}\) is generated using the \("generateSalt"\) method of the same library.
 
In particular, consider this 
scenario to obtain the Key for encrypting any input string:
Bob has an identifying key K, which is also shared with Alice, that he can identify himself with. Only both of them know this key K. Bob then encrypts \( \mathnormal({N_{1} || K}) \) using Alice's public key, and Alice decrypts it using its private Key to obtain \( \mathnormal{N_{1}}\) and K. Alice uses HMAC SHA-256 with \( \mathnormal({K || N_{1}}) \) to yield K(e) of 256 bit and HMAC SHA-256 with \( \mathnormal({K || N_{1}+1}) \) to yield K(m) of 256 bits. K(e) is the encryption key that is used for encrypting the message. K(m) is MAC Key that is used for creating a message authentication code (MAC) to ensure the integrity and authenticity of the message.
\\
To send a message to Bob, Alice needs to perform the following actions:
\begin{itemize}
    \item A new random 128-bit Initialization Vector (IV) is created.
    \item The message is encrypted using IV and K(e) as the Key.
    \item A SHA-256 HMAC is created with K(m) as Key and \( \mathnormal({IV || Encrypted message}) \) as data.
    \item Then she finally sends \( \mathnormal({IV || HMAC || Ciphertext}) \) to Bob
\end{itemize}

\begin{figure*}
\centering
\includegraphics[width=12cm, height=4cm]{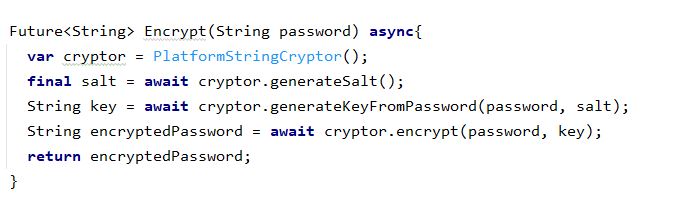}
\caption{Encryption prototype}
\label{fig:prototype}
\end{figure*}

The "generateKeyFromPassword" function from Flutter is used to generate the key using the method described above, where user-provided input is plain text and salt is IV. 
This generated key is used to encrypt any given input string (password etc) to obtain a secure token (AID/Ks) for our authentication model as shown in Figure ~\ref{fig: encrypt mac} and the prototype is shown in Figure ~\ref{fig:prototype}.
\section{IMPLEMENTATION AND TEST SETUP}
\subsection{Test Setup}
The following software and hardware were used for implementation and testing:

\begin{itemize}
    \item A LG G8X ThinQ smartphone with Qualcomm SM8150 Snapdragon Chipset, 1.0 GHz Octa-core (1x2.84 GHz Kryo 485 \& 3x2.42 GHz Kryo 485 \& 4x1.78 GHz Kryo 485) CPU, Adreno 640 GPU, 128GB 6GB RAM, and Android 9.0 (Pie) operating system.
    \item The PC used for implementing the Android application was a Windows 10 machine with Chrome Browser 90.0.4430.212 Version and an Intel(R) CoreTM i5-1035G1 CPU @ 1.00 GHz 8.00 GB of RAM.
    \item For the testing registration phase, the smartphone login phase was hosted on a local PC running Windows 10 OS on an Intel(R) CoreTM i5-1035G1 CPU @ 1.00 GHz with 8.00 GB of RAM. The API scripts were written in PHP and MYSQL and were hosted on the XAMPP server.
    \item The first device hosting the application and the second device used for login were on the same LAN during the testing.
    \item Though the Android App has been used to implement the demo, the proposal can be implemented in other browsers and mobile operating systems.
\end{itemize}

\subsection{Server Architecture}
The primary process to implement the proposed model is the authenticator web server. A web application server is required to implement the verifier system of the authentication protocol. The web application server communicates with both the first and second devices via REST API. REST (Representational State Transfer) is a set of architectural constraints, not a protocol or standard for implementing any web services. The proposed design approach lets the web application server communicate with both devices irrespective of their underlying operating system. 
The "PhpMyAdmin Xampp" was used to implement the web server application. The Android project was deployed and tested locally on a server, pre-installed with MYSQL database, Apache web server, Perl, and PHP to build an offline application with the desired functionality.

\subsection{Database Design}
Another vital component of this protocol design is the device vetting process by using a Bluetooth address. To assess both the devices, user and device identification information was required. Considering that, a primary database prototype was designed and also implemented. An email address and FbURL were used to identify any user in real time. Also, both the first and second device was authenticated using their Bluetooth hardware address against the registered user email address. The devices were associated for the first time during the login phase, and their Bluetooth addresses were updated for the first time in the database during this phase only. The MySQL and Flutter Real-time databases were both used for data storage and data persistence. The Firebase Realtime Database is cloud-hosted. Data is stored as JSON and synchronized in real time to every connected client. The Firebase database was used to store the user flutter auth ID, email address, and FbURL of the registered user. Subsequently, user details corresponding to this flutter auth ID are stored in the MYSQL database, ensuring a double layer of security.

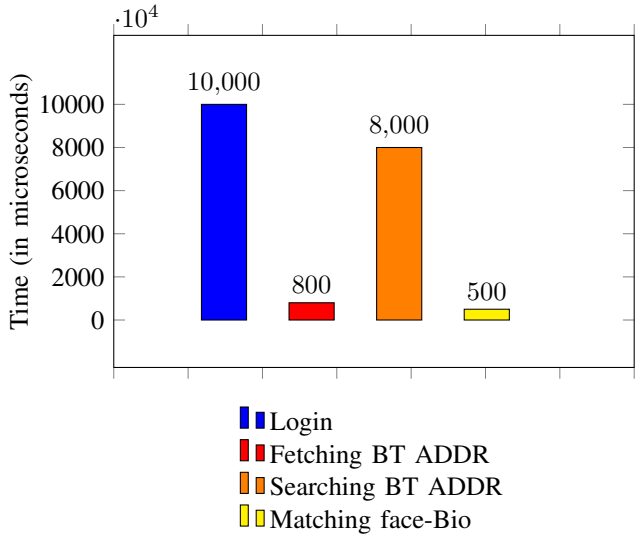
\begin{figure}
    \centering
    \begin{tikzpicture}
        \begin{axis}[
            height=6cm,
            width=8.5cm,
            ybar,
            bar width=0.6cm, 
            enlargelimits=0.2, 
            ylabel={Time (in microseconds)}, 
            xticklabels=\empty, 
            nodes near coords, 
            nodes near coords align={vertical}, 
            legend style={
                at={(0.5,-0.1)}, 
                anchor=north, 
                cells={anchor=west}, 
                draw=none 
            },
            xmin=-1, 
            xmax=4, 
            ymin=0, 
            ymax=11000, 
            ytick={0,2000,4000,6000,8000,10000}, 
            yticklabels={0, 2000, 4000, 6000, 8000, 10000}, 
        ]
        \addplot[fill=blue] coordinates {(0.5, 10000)};
        \addplot[fill=red] coordinates {(1, 800)};
        \addplot[fill=orange] coordinates {(1.5, 8000)};
        \addplot[fill=yellow] coordinates {(2, 500)};
        
        \legend{
           Login,
           Fetching BT ADDR,
           Searching BT ADDR,
           Matching face-Bio
        }
        \end{axis}
    \end{tikzpicture}
    \caption{Time needs through FlyBuy}
    \label{fig:Time-needs-through-FlyBuy}
\end{figure}

\begin{figure}
    \centering
    \begin{tikzpicture}
        \begin{axis}[
            height=6cm,
            width=8.5cm,
            ybar, 
            bar width=0.5cm, 
            enlargelimits=0.15, 
            ylabel={Time (in microseconds)}, 
            xticklabels={},
            nodes near coords, 
            nodes near coords align={vertical}, 
            legend style={
                at={(0.5,-0.1)}, 
                anchor=north, 
                cells={anchor=west}, 
                draw=none 
            },
            xmin=-1, 
            xmax=4, 
        ]
        \addplot[fill=blue] coordinates {(0, 4)};
        \addplot[fill=red] coordinates {(0.5, 13)};
        \addplot[fill=orange] coordinates {(1, 23)};
        \addplot[fill=yellow] coordinates {(1.5, 3)};
        \addplot[fill=green] coordinates {(2, 2)};
        \addplot[fill=pink] coordinates {(2.5, 1)};
        
        \legend{
            QR code scanning,
            OTP delivery to phone,
            Extra time for graphical password entry,
            Receiving push notifications,
            Connecting paired Bluetooth devices,
            Extra time due to NFC tap
        }
        \end{axis}
    \end{tikzpicture}
    \caption{Task Duration Comparison (in microseconds)}
    \label{fig:task-duration-comparison}
\end{figure}
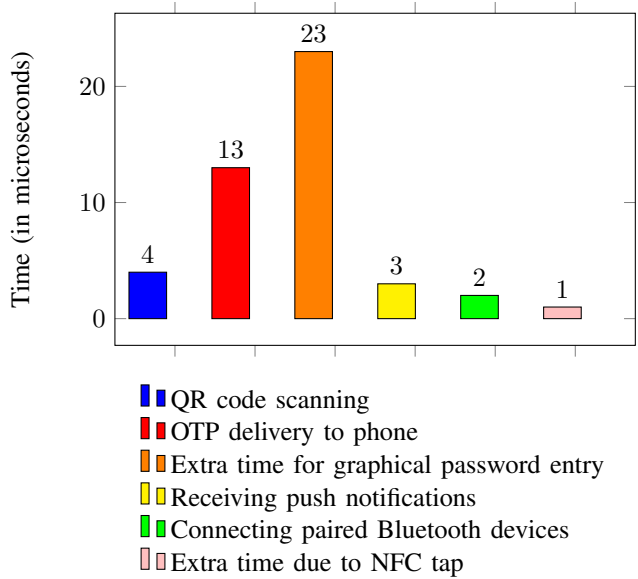

\subsection{Performance evaluation}
The performance of our proposed model was gauged in terms of the time taken for its various operations and the Memory and CPU utilization. "Another Monitor" Android App was deployed over both the devices to record the Memory and CPU utilization  of our Android App "FlyBuy". The minimum CPU utilization of the FlyBuy Android App is 0.28\% on both devices, and the maximum CPU utilization was 4.80\% on the both device. The memory utilization for the first device was 12.2 MB and for the second device 13.1 MB. The statistics confirm that the Memory and CPU utilization of our model is low enough, and it could be used for everyday login purposes. The memory utilization, CPU utilization, and login time of our model have not been compared with the other existing schemes because of the underlying reasons outlined below:
\begin{itemize}
    \item The application login time is decided by two factors, i.e., the speed and availability of the web application server. The login time was captured when only one user tried to log in to the web server as shown in Fig.~\ref{fig:Time-needs-through-FlyBuy} Since it is not feasible to record login time on a commercial website using other authentication schemes (User-PWD, Push notification-based, QR Code) because they used to manage multiple customers over the same period.
    \item The Memory and CPU Utilization depends on the type of application supported by the App. For example, an online social media app such as Instagram App consumes a lot of Memory and CPU because of the high media content they receive and process. Thus, a direct comparison of CPU and Memory Utilization is not a fair idea.
    \item Some of the non-commercial proposed schemes haven't mentioned anything related to  Memory, CPU, and login time in the past.
\end{itemize}
We did numerous trials to record the additional average time for two parameters (1) when the first device tried to connect via Bluetooth technology with the second device and (2) when the first device tapped against the second device to transfer the token. Approximately 100 experiments were carried out, as shown in Figure ~\ref{fig:task-duration-comparison}
The average time values for the push notification, an extra time for graphical password entry, OTP delivery to a phone, and QR code scanning has been obtained from \cite{push1} respectively. Our experiment shows that the average time required for the first device to tap against the second device and connect two devices via Bluetooth technology takes less time than the schemes mentioned above.

\section{Security analysis}
This section discusses the security of the proposed scheme against
the attacks which can be carried out by an attacker to compromise
the authentication scheme or steal the user credentials.
\subsection{\textbf{Security against RT MITM phishing}}
Device-specific Bluetooth address \( \mathnormal{BT_{1}} \) and \( \mathnormal{BT_{2}} \) and user-specific \( \mathnormal{AID} \) cannot be acquired through remote desktop monitoring / remote screen relaying, malicious browser extensions or on phishing websites and hence cannot be relayed to an authentic website in real time to cause an RT MITM attack or a CR MITM phishing attack. The attacker will not be able to acquire \( \mathnormal{BT_{1}} \) because this is automatically updated by the first device to the database server and the second device detects the existence of \( \mathnormal{BT_{1}} \) in it's proximity once update is done. Also, an attacker can't obtain the concatenated authentication string which is generated in encrypted form (encryption key: legitimate Android app private key concatenated with server SALT) in the local storage of the authentic Android App and transferred to the second device through NFC tap and deleted once transfer is done. The second device also deletes this secret concatenated authentication string once the login is completed or if the timer started by the server expires during the inactive session. Also, the attacker will need the \( \mathnormal{E_{SK+SALT}(AID)} \) stored on the registered smartphone Android App.
\subsection{\textbf{Security against CR MITM phishing}}
CR MITM phishing is not possible because the use of \( \mathnormal{BT_{1}} \) as the first device identification token and \( \mathnormal{BT_{2}} \) as the second device identification token since it is impossible to access the Bluetooth address of the first device connected to the user second device via Web Bluetooth APIs as Web Bluetooth APIs can only access the BLE devices physically connected to the attacker’s PC. 
\subsection{\textbf{Malicious Browser Extension based attacks}}
As user, does not type any information on the website the credentials stealing attacks which happens through keystroke logging, PWD, and HTML form data sniffing cannot happen. 
\subsection{\textbf{Host malware based keystroke logging}}
Keystroke logging-based attacks that steal credentials entered 
by the user over websites can also be avoided with the use of 
Bluetooth address and NFC as the device identification token as \( \mathnormal{BT_{1}} \) is
automatically retrieved by the websites and the secret is transferred to the second device through an NFC tap and is automatically retrieved by servers to verify the user's biometrics. 
\begin{table*}[h]
\caption{Model Checker Result}
\label{tab:avispa}
\begin{tabularx}{\textwidth} { 
  | >{\raggedright\arraybackslash}X 
  | >{\raggedright\arraybackslash}X
  | >{\raggedright\arraybackslash}X | }
 \hline
\textbf{Backend} & \textbf{Summary} & \textbf{Statistics} \\
\hline
OFMC & SAFE & parse time: 0.00s  searchTime: 35.55s  visitedNodes: 19136 nodes  depth: 15 piles \\
\hline
CL-Atse & SAFE & Analysed: 309673 states  Reachable: 309673 states  Translation: 0.01 seconds  Computation: 31.05 seconds  \\
\hline
 \end{tabularx}
\end{table*}


\section{EXPERIMENTAL RESULTS \& DISCUSSIONS}
\begin{figure*}[!htb]
    \centering
    \begin{minipage}[t]{0.5\textwidth}
        \centering
        \includegraphics[width=4cm, height=8cm]{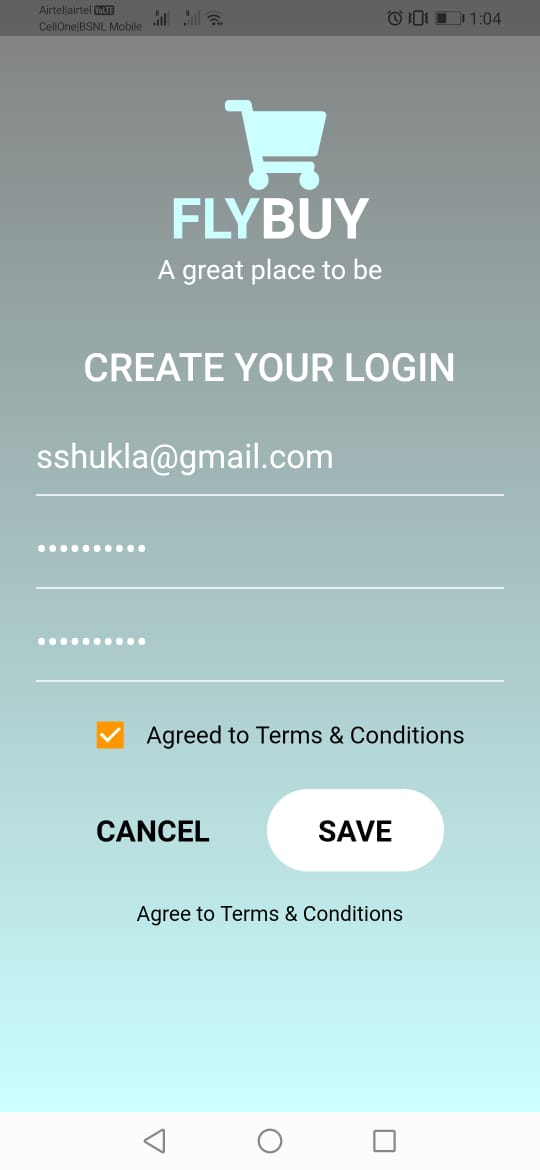}
        \caption{FlyBuy account creation screen}
        \label{fig:sc1}
    \end{minipage}%
    \begin{minipage}[t]{0.5\textwidth}
        \centering
        \includegraphics[width=4cm, height=8cm]{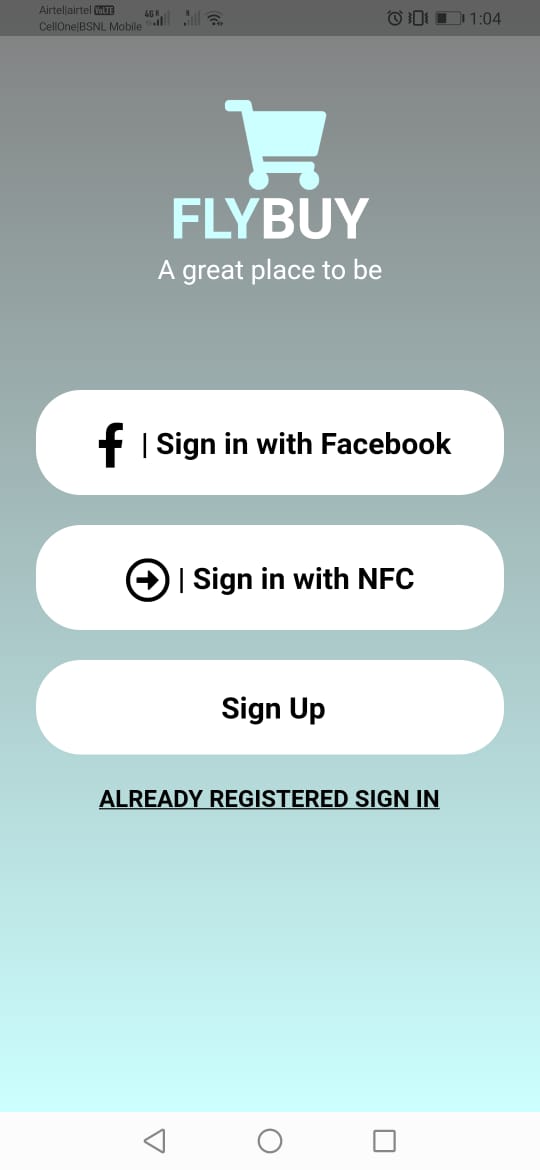}
        \caption{FlyBuy home page screen with sign-in button}
        \label{fig:sc2}
    \end{minipage}
    
    \vspace{0.5cm} 
    
    \begin{minipage}[t]{0.5\textwidth}
        \centering
        \includegraphics[width=5cm, height=10cm]{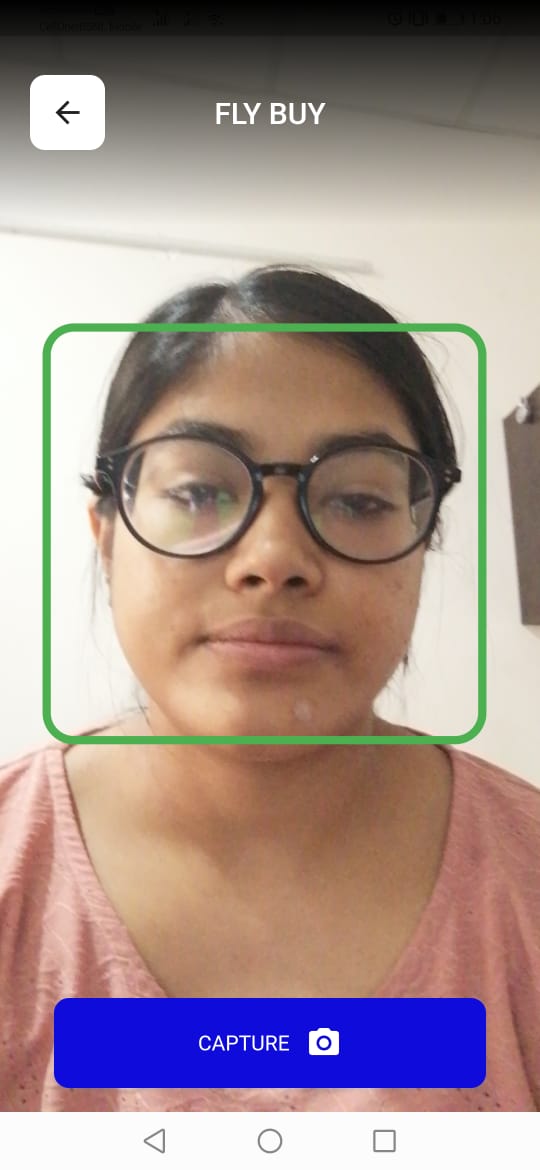}
        \caption{User taking their face biometric in real time using FlyBuy app}
        \label{fig:sc3}
    \end{minipage}%
    \begin{minipage}[t]{0.5\textwidth}
        \centering
        \includegraphics[width=5cm, height=10cm]{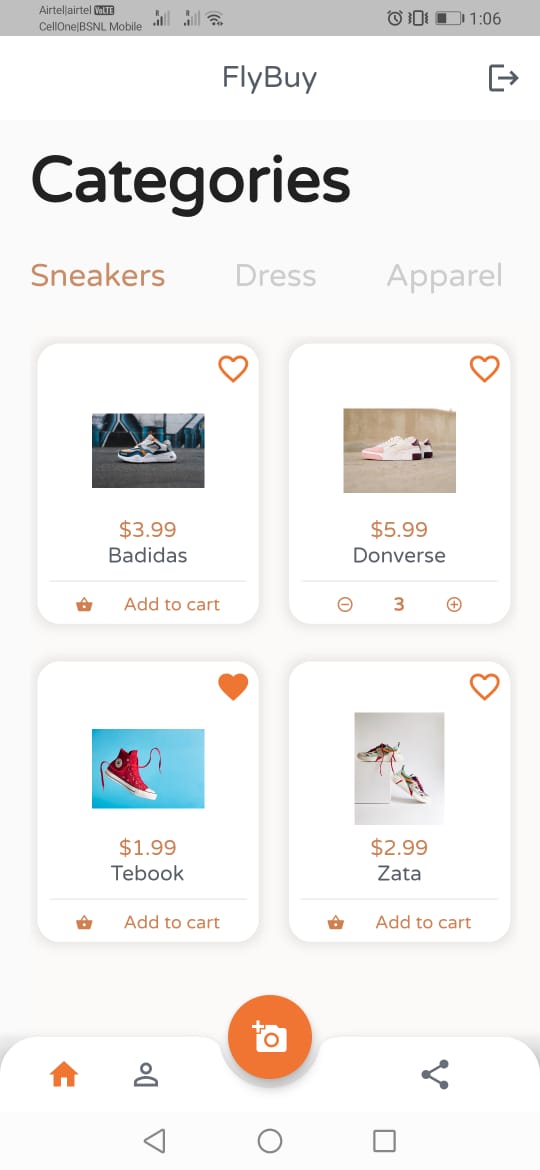}
        \caption{FlyBuy logged-in screen after the user authenticates on the second device}
        \label{fig:sc4}
    \end{minipage}
\end{figure*}

This section highlights the modeling and verification of the authentication protocol and above discussed authentication systems are
compared in terms of tokens used by the system, the number
of tokens that need to be remembered, the number of taps
required to send a token, and additional requirements. This section also draws a comparison between the existing authentication schemes in terms of usability, deployability, and security using the Bonneau et al. framework.
\\
Fig ~\ref{fig:sc1} shows the registration screen using which the user will create an account. Then the user can sign in user sign-in screen as shown in Fig ~\ref{fig:sc2}. Once user has logged in on the first device they can use the first device to login in the second device by taking pictures using the camera as shown in Fig ~\ref{fig:sc3}. Once the image is captured a secret file is generated that is transferred to the second device using NFC tap, after which the second device will search for the first device's proximity and allow the user to log into the second device as shown in Fig ~\ref{fig:sc4}

\subsection{Model Checking: Overview \& Results}
\subsubsection{Overview}
In order to verify that the web authentication protocol assures the secrecy and authenticity of communication between devices and the web server, a model checker must be used. Therefore, AVISPA \cite{report23} model checker was used to inspect secrecy and authenticity properties. The Automated Validation of Internet Security Protocols and Applications (AVISPA) tool provides a suite of applications for building and analyzing formal models of security protocols \cite{report23} for large-scale security protocols. The protocol models are written in the High-Level Protocol Specification Language or HLPSL. The AVISPA Tool comprises four backends for backend verification \cite{report24}:
\begin{itemize}
    \item OFMC (On-the-fly model checker)
    \item CL-AtSe (Constraint Logic-based Attack Searcher)
    \item SATMC (SAT-based Model-Checker)
    \item TA4SP (Tree-based model checker)
\end{itemize}
OFMC's backend comes in handy for detecting, guessing, and carrying out replay attacks. The SATMC and CL-Atse backend platforms are generally used to check the bounded number of protocol falsifications and sessions. TA4SP backend furnishes unbounded security protocol verification using tree-based languages \cite{report23}.
T represents a server, Bob represents the first device, and Alice represents the second device for the proposed authentication model. Kat, Kbt, and Kit are the keys commonly known between the second device and server, the first device and server, and the intruder and the server, respectively.
In AVISPA, there exist two security goals. In order to verify if the devices are authenticated to the server and to each other, and the first device Bluetooth address, the following goals were outlined:
\begin{enumerate}
    \item Authentication on the first device
    \item Authentication on the second device
    \item Authentication on the webserver
\end{enumerate}
Furthermore, in order to check if the communication was kept secret, the following goal was outlined:
\begin{itemize}
    \item Secrecy of secret key (Ks) (used for all token/data encryption)
\end{itemize}

\subsubsection{Results}
The Lenovo laptop was used during our experiments. The
laptop computer is a Windows 10, which has 8.00 GB RAM, a 1.00 GHz Intel(R) CoreTM i5-1035G1 processor.
The verification results are summarized in Table ~\ref{tab:avispa}. The CL-AtSe backend platform was used to verify the bounded number of sessions. CL-Atse completed the protocol verification in 31.05 seconds by analyzing 309673 states. As a result, this backend doesn't find any attack on the proposed protocol. In order to find replay and guessing attacks, the proposed model was verified by the OFMC backend. A heuristic search algorithm was run by OFMC with 15 piles and analyzed a total of 19136 nodes, and it was found SAFE from any of the possible attacks.
\begin{figure*}
\centering
\includegraphics[width=18cm, height=12cm]{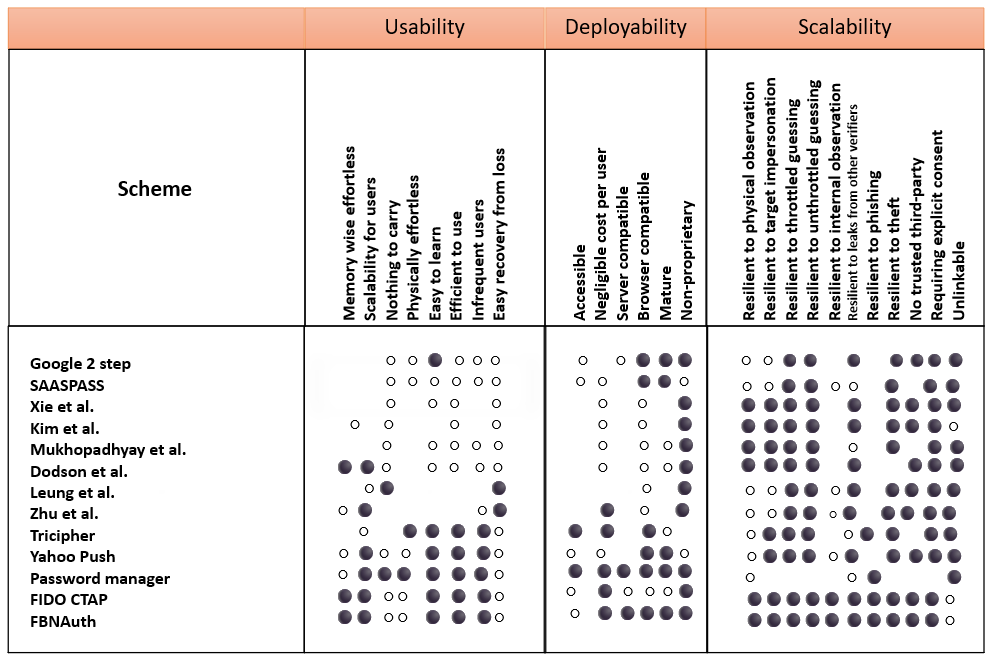}
\caption{Scheme Evaluation \textbullet\ indicates that the scheme fully carries the characteristic, \textopenbullet\ indicates that the scheme partially carries the characteristic (the Quasi prefix).}

\label{fig:bonneau}
\end{figure*}

\subsection {Comparison: Usability}
Above discussed authentication systems are compared in terms of tokens used by the system, the number of tokens that need to be remembered, the number of taps required to send a token, and additional requirements such as the need for Internet on a smartphone for the model to work for Smartphone (second device) login. After comparison, it was found that most of the schemes only need a smartphone as an additional requirement which most Internet users carry every day. Other schemes need driver modules to be installed on a PC, hardware token, GPS, trusted third party, etc. The requirement to have data or a Wi-Fi connection in the smartphone while logging in from a PC is a crucial factor in getting both the incurred cost using the proposed scheme and the issues that might arise due to the unavailability of internet connectivity on a smartphone. The comparative discussion has been outlined in Table ~\ref{tab:comp2}
\begin{table*}[h]
\caption{Comparison in terms of number of tokens used, number of tap,s and their security}
\label{tab:comp2}
\begin{tabularx}{\textwidth} { 
  | >{\raggedright\arraybackslash}X 
  | >{\raggedright\arraybackslash}X
  | >{\raggedright\arraybackslash}X
  | >{\raggedright\arraybackslash}X
  | >{\raggedright\arraybackslash}X
  | >{\raggedright\arraybackslash}X | }
 \hline
\textbf{Scheme} & \textbf{Tokens used by scheme} &\textbf{Number of touch/tap} &\textbf{Token to be remembered by user} &\textbf{Additional Needs} &\textbf{The need of internet on phone} \\
\hline
Google 2 Step \cite{report5} & 3 - U, PWD, OTP on SP & 3-U,PWD, SUB & 2-U,PWD & Cellphone & N \\
\hline
SAASPASS \cite{report19} & 3 - U, PWD, OTP on App & 3 - U,PWD, OTP,SUB & 2-U,PWD & smartphone & Y \\
\hline
Xie et al. \cite{rt3} & 4 - U, PWD, DH, Private Up  & 4-U,PWD, PC Cam,SUB  & 2-U,PWD & PC Cam, Smartphone & N \\
\hline
Kim et al. \cite{rt1} & 4 - U, PWD, Session ID, Secret Key & 3-U,PWD, SUB & 2-U,PWD & Smartphone with GPS & Y \\
\hline
Mukhopadhyay et al. \cite{mukho} & 3 - U, PWD, Secret key in SP & 3-U,PWD, SUB & 2-U,PWD & Smartphone, TTP & Y \\
\hline
Dodson et al. \cite{dodson} & 4 - U, PWD, Secret key, QR code & 4-U,PWD, QR code, SUB & 0 - NIL & Smartphone & Y \\
\hline
Leung et al. \cite{rt2} & 4 - U, PWD, Secret key, OTP CAPTCHA & 4-U,PWD, CAPTCHA, SUB & 2-U,PWD & NIL & NA \\
\hline
Zhu et al. \cite{zhu} & 3 - U, SALT, PWD CAPTCHA & 3-U,PWD CAPTCHA,SUB& 2-U,PWD & NIL & NA \\
\hline
Tricipher \cite{report20} & 4 - U, PWD, TPM Secret key, TACS credential & 4-U,PWD, Secret token, SUB & 2-U,PWD & CAPI driver, Separate hardware, TPM & N \\
\hline
RSA SecurID Token \cite{report22} & 4 - U, PWD, HW token, PIN & 4-U,PWD, Secret token, SUB & 2-U,PWD & Separate hardware & NA \\
\hline
Yubikey U2F \cite{report12} & 5 - KPUB, KPRIV, Counter, U, PWD & 4-U,PWD, Secret Key, SUB & 2-U,PWD & Separate hardware & NA \\
\hline
Push login \cite{push1} & 3 - U, PWD, SP & 4-U,PWD, Push notification, SUB & 1-U & Smartphone & Y \\
\hline
Password Managers \cite{report7} & 3 - U, PWD, master key  & 1 - SUB & 1 - Master PWD & NIL & NA \\
\hline
U-PWD & 2 - U, PWD & 3- U,PWD, SUB & 2-U,PWD & NIL & NA \\
\hline
Proposed scheme & 4 - BT1, BT2, AID, Face-Bio & 3-Face-Bio, NFC tap, PC tap & 0 - NIL (Face scan) & Smartphone & N \\
\hline
\end{tabularx}
\end{table*}

\subsection{Assessment with Bonneau et al. Framework}
This subsection draws a comparison between the existing authentication schemes in terms of usability, deployability, and security using the Bonneau et al. framework.  
In the comparison table, some of the values are directly taken from the study \cite{study}, and some other values are modified according to the analysis performed on the scheme against the latest cyber-attacks. A brief description of different Bonneau et al.'s framework parameters has been provided, and benefits offered by existing schemes have been discussed.
\subsubsection{Usability}
\begin{itemize}
    \item  \textbf{Memory wise effortless} Nearly all the existing schemes are not memory-wise effortless due to the fact that they want the user to remember username and password. Yahoo Push login \cite{push1}, Password managers \cite{report7}, Kim et al. \cite{rt1} reduces the number of tokens to be memorized and input entered by the user and hence are categorized as scheme offering partial benefit while our proposed scheme and Dodson et al.'s authentication scheme \cite{dodson} doesn't require the user to enter any input on the website are considered as "offering the benefit" scheme.
.

    \item \textbf{Scalable for users} 
    If logging over multiple web accounts using a similar scheme increases the load on the user, then that authentication scheme is not a scalable one. Our scheme offers this benefit where the user needs to just connect the BT mobile with the PC once and have to log into their Smartphone App before in order to log into their websites on PC.
    \item \textbf{Nothing and quasi-nothing to carry} 
     It is considered that a user carrying his/her smartphone is almost equivalent to carrying nothing. Our scheme offers this benefit given the fact that nearly every internet user carries a smartphone. Rather if the user is required to carry an additional device or hardware, then the scheme doesn't offer this benefit. For example, Tricipher \cite{report20} login is client-system dependent, due to which it doesn't offer this benefit.
    \item \textbf{Physically effortless} 
     If the user doesn't need to perform extra efforts such as tapping a button or entering credentials, then the scheme is considered to be physically effortless. There are so many existing schemes that do not offer this benefit completely, as the user might have to fetch a PIN or OTP in order to interact with the smartphone. Some schemes require the user to perform more interaction with the device, such as scanning Barcode or CAPTCHA entry. Our proposed model is rationally physical effortless since NFC tap, Bluetooth pairing, or capturing Images using a Camera is a one-click process.
    \item \textbf{Easy to learn} 
    Any scheme which a user can learn with the basic knowledge of computer and uses the scheme, then that scheme is considered to be easy to learn. The schemes such as CAPTCHA or graphical passwords, and QR code scanning don't offer this benefit mostly. However, the proposed scheme is easy to understand and learn since the user only needs to know how to do BT pairing of the devices. Also, BT pairing is done once between two devices by means of a one-click process. Even people belonging to the old age group lacking technical knowledge can be guided once to perform BT pairing, and after that, they can simply connect with a PC at any time of the day. 
    \item \textbf{Efficient to use}  
    If the login time is minuscule, then the authentication scheme is considered to be an efficient one to be used.
    The existing schemes such as QR code-based schemes OTP based schemes offer this benefit partially only, while schemes such as graphical password do not offer this benefit at all due to the time it takes to generate and show CAPTCHAs, other than the time taken by any user to enter it using mouse clicks. Our proposed model is efficient to use due to two facts: (1) the time required for pairing Bluetooth devices (2) tapping using NFC to transfer tokens takes on average less than the time required to receive OTP or scan any Barcode. Apart from this, schemes such as QR codes and graphical password requires user understanding for scanning or entering the credentials on the website.
    \item \textbf{Infrequent error} 
    Several schemes don't offer this benefit. The user might not always obtain a valid SMS or OTP in a stipulated time. Schemes such as QR code / barcode-based schemes have frequent errors since they require the user's to match geolocations based on IP address or perform multiple Barcode scans in some situations. In our proposed scheme, there is very less or no possibility of error, other than the BT address pairing problem, which is rare to occur during the login phase.
    \item \textbf{Easy recovery} 
    If a user account can be easily recovered even after the loss of a legitimate device or secret token, then it is said to offer this benefit. Several authentication schemes innately provide this easy recovery benefit, but few schemes that use hardware tokens or smartphones might require additional effort for account recovery.
\end{itemize}

    \subsubsection{Deployability}
    \begin{itemize}
    
        \item \textbf{Accessible} 
    The authentication schemes such as moving CAPTCHAs or Barcode/QR code-based schemes use a graphical password or visual password, and hence they are inaccessible to visually or physically impaired users. Such an audience can leverage to touch-based login system. Even if the authentication scheme requires users to gain some technical knowledge before operating their smartphone, then also scheme is considered to be not offering this benefit.
    \item \textbf{Negligible cost per user} 
    If an authentication scheme requires a separate hardware token or OTP, then it doesn't offer this benefit. However, if it requires having Wifi or Cellular mobile data on a smartphone to complete the authentication process, then it might be considered to be offering this benefit since internet connection is generally available on the smartphone.
    \item \textbf{Server Compatible} 
    Several authentication schemes are server-compatible if they don't require a separate mechanism in order to complete the authentication process, such as the generation of a certificate or QR code generation or scanning using an Android App. In other scenarios where server implementation is platform-dependent or requires specific floating CAPTCHAs appearance, then the scheme doesn't offer server-compatible benefits. Our proposed model and Google 2-step authenticator require minuscule changes at the server end, such as obtaining Bluetooth address or OTP generation. However, traditional username and password-based schemes or password managers completely offer this benefit to the end-user.
    \item \textbf{Browser compatible} 
    If a user needs to install specific modules or software for their browsers or needs to change their browsers for the authentication scheme to work, then the scheme doesn't offer this benefit. Any scheme providing apps and extensions without the requirement of extra support by browsers such as different versions of scripting language or HTML form-related versions, then these schemes are said to offer partial benefit to the user.
    \item \textbf{Mature} 
    Those schemes that have been rigorously tested and widely adopted by the public are considered to be mature. Those authentication schemes, which are incremental addition to the existing ones, are said to offer partial benefits. Our proposed work also has minuscule verifiable design changes such as obtaining Bluetooth addresses and capturing real-time face and AID during login from existing MFA schemes.
    \item \textbf{Non-proprietary} 
    If the scheme requires explicit approval from the developer for its use are said to be proprietary and offers no benefit.
    \end{itemize}
    \subsubsection{Security}
    \begin{itemize}
 
        \item \textbf{Resilient to physical observation} 
    Any scheme in which all the user credentials cannot be captured by physical observation, such as thermal imaging of the keyboard or shoulder surfing or keyboard filming, etc., is said to be resilient to physical observation. Other schemes such as a graphical password or Google 2-step authenticator-based schemes etc., offer this partially as this technique can be used to capture all the user credentials. However, schemes like the Push notification-based scheme, QR code-based scheme, and our proposed work are secure from all such attacks as credentials cannot be obtained by physical observation.
    \item \textbf{Target impersonation} 
    Most of the schemes use either a QR code, secret key, hardware token, or OTP for login purposes which makes it difficult for an attacker with basic knowledge of user details such as age, name, date of birth, etc., to compromise the system and get the user's account access.
    \item \textbf{Throttled and unthrottled guessing} 
    Only traditional username and password-based schemes are vulnerable to password breaking via throttled or unthrottled guessing because of the presence of QR code or graphical password or hardware tokens, which are not easy to guess by an attacker.
    \item \textbf{Internal observation} 
    The internal observation can be carried out by sniffing client and server communication using MBE or by host keylogging. Most of the schemes are unsafe from this attack.
    \item \textbf{Resilient to leaks from other verifiers} 
    Those schemes that involve a third party during the login phase do not offer this benefit.
    \item \textbf{Phishing} 
    The authentication schemes that are vulnerable to traditional or RT/CR MITM are also vulnerable to phishing (see Table ~\ref{tab:comp2})
    \item \textbf{Resilient to theft} 
    Nearly every authentication scheme is resilient to the theft of a smartphone or the hardware token except password manager-based schemes and Dodson et al. authentication schemes, in which the attacker is capable of directly accessing a user account without providing any user information such as PIN on the website if they are in access of the smartphone or PC. 
    \item \textbf{No trusted third party} 
    The authentication scheme that doesn't use a trusted third party during the login process provides this benefit. SAASPASS \cite{report18}, Mukhopadhyay et al. \cite{mukho}, Tricipher \cite{report20} and password manager uses a trusted third party for login purposes.
    \item \textbf{Explicit consent} 
    Nearly every authentication scheme requires the user to explicitly provide their consent during the login phase, except password managers since they used to auto-fill the user information. 
    \item \textbf{Unlinkable} 
    Several schemes offer this benefit and are not used to link with any user. The schemes such as Kim et al.'s \cite{rt1} are linkable because they use an IP address and this IP address is used to verify the same user identity who is trying to perform login to various websites. Our proposed scheme is somewhat linkable due to the fact that a similar BT device is used to perform login to various websites and hence doesn't offers this benefit. 
    \end{itemize}
    Our analysis shows that the proposed model performs better in comparison to the other existing schemes, as shown in Figure ~\ref{fig:bonneau}.

\section{CONCLUSION AND FUTURE WORK}
This paper presents a robust and secure multi-factor passwordless authentication scheme that is capable of handling phishing attacks such as RT MITM, CR MITM phishing, MBE-based attacks, and App spoofing against the above existing schemes \cite{c1,c2,c3,c4,c5,c6,c7,c8,c9} \cite{rt1,rt2,rt3,rt4} 
\cite{mbe1},\cite{mbe2},\cite{mbe3},\cite{mbe4},\cite{mbe5},\cite{mbe6},\cite{mbe7},\cite{mbe8}
\cite{report1,report2,report3,report4,report5,report6,report7,report8,report9,report10,report11,report12,report13,report14,report16,report17,report18,report20}. The proposed scheme has certain advantages outlined below:
\begin{enumerate}
    \item The number of credentials that the user needs to remember and enter is reduced to zero.
    The user doesn't need to remember any token while performing authentication, while other credentials are obtained automatically with the explicit consent of the user, such as capturing face biometric, Bluetooth address (after user enables the Bluetooth functionality on the device), and token transfer using NFC tap (after user enables NFC functionality on the device). This feature ensures security against RT MITM attacks since an attacker cannot obtain all the user credentials and relay them in real time for authentication on a legitimate website.
    \item It is not feasible for MBE or any malicious peer App to sniff data or log any user information since the user is never going to provide any credentials on the App and also due to same-origin policy, thus avoiding MBE-based phishing attacks.
    \item The CR MITM and App spoofing attacks were avoided by the use of Bluetooth address and AID.
\end{enumerate}
Another advantage of the proposed model is that it's not client-side dependent, unlike other schemes such as Tricipher. The model was implemented and tested to analyze its efficiency in terms of memory and CPU utilization. The results obtained showed satisfactory performance. Also, a comparison was carried out in terms of usability, deployability, and security against existing schemes, which shows that it performed better than others and ensured the same level of security in comparison with other schemes. In our proposed work, a single Android App was used to log in to the website. This Android App will be provided to the user by a trusted third party (i.e., the organization). Every legitimate website has to fetch, verify and store the Bluetooth address of smartphones of their registered users during the login phase as part of the proposed work. On comparison with FIDO CTAP, Our method achieve the same goal but employ different approaches. In our method, we utilize face biometrics as an authentication token. In contrast, FIDO CTAP depends on a key stored either on a smartphone or hardware security key, and it uses face biometrics and fingerprints for user verification during cryptographic authentication token transfer. FIDO CTAP's reliance on established protocols
and challenge-response mechanisms further strengthens its security posture. We presents a fresh perspective on secure authentication, leveraging readily available devices and prioritizing face biometrics. FIDO CTAP is not server-compatible due to its reliance on specialized hardware or biometric sensors for authentication, which introduces complexity and platform dependencies in server implementation. It requires specific devices or biometric data processing, making it less accessible and straightforward for end-users and potentially challenging to implement uniformly across different environments. Therefore, it does not align with the concept of server compatibility as defined, which emphasizes authentication schemes that don't require additional mechanisms or platform-specific implementations while our scheme depends only on BLE and NFC-enabled mobile phones. In the case of browser compatability FIDO CTAP requires FIDO-enabled browser so it is not compatible with every browser while the scheme is compatible with every browser. When talking about maturity FIDO CTAP is also less mature due to its heavy reliance on specific hardware devices for authentication. The dependency on these devices introduces potential single points of failure and limited backup mechanisms compared to more established software-based authentication approaches.

However, our proposed scheme has some current limitations, as discussed below:
\begin{itemize}
    \item The authentication scheme requires that the user must possess a smartphone when he/she is trying to authenticate to a website using his/her PC. This was reported by a survey \cite{report25} that the number of users present across the globe is 6055 billion until 2020, and there are likely chances that this number will increase to 7516 billion by 2026. One of the surveys has mentioned that 4.28 billion people use the internet indicating 90\% of the worldwide internet population uses smartphone to go online \cite{report26}. Also, two billion NFC-enabled devices like a smartphone are in use today (IHS). In other words, 20\%+ of the world’s population have access to NFC. This growth will be spurred, in part, by seven billion smartphone subscriptions by 2022 with NFC as a key enabler because it provides consumers the necessary ease of use and convenience making it a useful, universal technology. From the above discussion, it can be assumed that any user using internet services and having technology similar to NFC and BLE-enabled smartphones can also use our proposed scheme for authentication purposes.
    
    \item The proposed scheme has not been analyzed in terms of security against host-based malware. The malware capabilities include capturing keystrokes of the user, accessing data, resources, and system files, etc. Some malware, such as keystroke loggers, are capable of breaking most of the authentication schemes, such as OTP/PIN-based schemes. Other malware such as screen loggers is capable of breaking QR codes and graphical password-based schemes and even obtaining the user credentials sent through separate hardware tokens via memory dumping and analysis. Malware that can access the master password or Windows password of the user from the system storage can break the Google password manager scheme by compromising user passwords stored inside Chrome and third-party password managers like LastPass \cite{report6} as well. Hence, this attack is not in the current scope of work, and can be considered as future work.
\end{itemize}

\textbf{Conflict of interest}: The authors have no Conflict of interest to declare
concerning this paper.


\begin{thebibliography}{00}





\bibitem{wiki1}
Wikipedia Phishing. Available: \url{https://en.wikipedia.org/wiki/Phishing}

\bibitem{wiki2}
Phishing.org, Phishing Techniques. Available: \url{https://www.phishing.org/phishing-techniques}

\bibitem{phish1}
Badra, Mohamad, Samer El-Sawda, and Ibrahim Hajjeh. "Phishing Attacks and Solutions." In \textit{Proceedings of the 3rd International Conference on Mobile Multimedia Communications}, pp. 42. ICST (Institute for Computer Sciences, Social-Informatics and Telecommunications Engineering), 2007.

\bibitem{phish2}
Dodge, Ronald C., Curtis Carver, and Aaron J. Ferguson. "Phishing for user security awareness." \textit{Computers \& Security} 26, no. 1 (2007): 73-80.

\bibitem{phish3}
Hong, Jason. "The State of Phishing Attacks." \textit{Commun. ACM} 55, no. 1 (2012): 74–81. Available: \url{https://doi.org/10.1145/2063176.2063197}

\bibitem{phish4}
Varshney, Gaurav, Manoj Misra, and Pradeep K. Atrey. "A phish detector using lightweight search features." \textit{Computers \& Security} 62 (2016): 213-228. Available: \url{https://doi.org/10.1016/j.cose.2016.08.003}


\bibitem{phish5}
Islam, Rafiqul, and Jemal Abawajy. "A multi-tier phishing detection and filtering approach." \textit{Journal of Network and Computer Applications} 36, no. 1 (2013): 324-335. Available: \url{https://doi.org/10.1016/j.jnca.2012.05.009}

\bibitem{phish6}
Ramesh, Gowtham, Jithendranath Gupta, and P.G. Gamya. "Identification of phishing webpages and its target domains by analyzing the feign relationship." \textit{Journal of Information Security and Applications} 35 (2017): 75-84. Available: \url{https://doi.org/10.1016/j.jisa.2017.06.001}

\bibitem{rt1}
Kim, Seung-Hyun, Daeseon Choi, Seung-Hun Jin, and Sung-Hoon Lee. "Geo-Location Based QR-Code Authentication Scheme to Defeat Active Real-Time Phishing Attack." In \textit{Proceedings of the 2013 ACM Workshop on Digital Identity Management}, pp. 51–62. Association for Computing Machinery, 2013. Available: \url{https://doi.org/10.1145/2517881.2517889}

\bibitem{rt2}
Leung, Chun-Ming. "Depress phishing by CAPTCHA with OTP." In \textit{2009 3rd International Conference on Anti-counterfeiting, Security, and Identification in Communication}, pp. 187-192. IEEE, 2009. Available: \url{https://doi.org/10.1109/ICASID.2009.5276926}

\bibitem{rt3}
Xie, Mengjun, Yanyan Li, Kenji Yoshigoe, Remzi Seker, and Jiang Bian. "CamAuth: Securing Web Authentication with Camera." In \textit{2015 IEEE 16th International Symposium on High Assurance Systems Engineering}, pp. 232-239. IEEE, 2015. Available: \url{https://doi.org/10.1109/HASE.2015.41}

\bibitem{rt4}
Lu, Yanrong, Lixiang Li, Haipeng Peng, and Yixian Yang. "An Energy Efficient Mutual Authentication and Key Agreement Scheme Preserving Anonymity for Wireless Sensor Networks." \textit{Sensors} 16, no. 6 (2016): 837. Available: \url{https://www.mdpi.com/1424-8220/16/6/837}

\bibitem{mbe1}
Dhawan, Mohan, and Vinod Ganapathy. "Analyzing Information Flow in JavaScript-Based Browser Extensions." In \textit{2009 Annual Computer Security Applications Conference}, pp. 382-391. IEEE, 2009. Available: \url{https://doi.org/10.1109/ACSAC.2009.43}

\bibitem{mbe2}
Saini, Anil, Manoj Singh Gaur, Vijay Laxmi, and Mauro Conti. "Colluding browser extension attack on user privacy and its implication for web browsers." \textit{Computers \& Security} 63 (2016): 14-28. Available: \url{https://doi.org/10.1016/j.cose.2016.09.003}

\bibitem{mbe3}
Saini, A., Gaur, M.S., Laxmi, V., Singhal, T., Conti, M. "Privacy Leakage Attacks in Browsers by Colluding Extensions." \textit{Information Systems Security. ICISS 2014.} (2014). Available: \url{https://doi.org/10.1007/978-3-319-13841-1_15}

\bibitem{mbe4}
Varshney, Gaurav, Manoj Misra, and Pradeep Atrey. "Browshing a new way of phishing using a malicious browser extension." In \textit{2017 Innovations in Power and Advanced Computing Technologies (i-PACT)}, pp. 1-5. IEEE, 2017. Available: \url{https://doi.org/10.1109/IPACT.2017.8245147}

\bibitem{mbe5}
Ocean Science Discussions. "Dynamically constrained ensemble perturbations – application to tides on the West Florida Shelf." Available: \url{https://os.copernicus.org/preprints/6/1/2009/osd-6-1-2009.pdf}

\bibitem{mbe6}
Fraser, Norman. "The usability of picture passwords perturbations – application to tides on the West Florida Shelf." Available: \url{https://www.tricerion.com/wp-content/uploads/2013/09/Usability-of-picture-passwords.pdf}

\bibitem{mbe7}
Dhamija, Rachna, and J. D. Tygar. "The Battle against Phishing: Dynamic Security Skins." In \textit{Proceedings of the 2005 Symposium on Usable Privacy and Security}, pp. 77–88. ACM, 2005. Available: \url{https://doi.org/10.1145/1073001.1073009}


\bibitem{mbe8}
Varshney, Gaurav, Manoj Misra, and Pradeep K. Atrey. "Detecting Spying and Fraud Browser Extensions: Short Paper." In \textit{Proceedings of the 2017 on Multimedia Privacy and Security}, pp. 45–52. Association for Computing Machinery, 2017. Available: \url{https://doi.org/10.1145/3137616.3137619}

\bibitem{report1}
APWG. "Phishing Activity Trends Report, 1st Quarter 2021." Available: \url{https://docs.apwg.org/reports/apwg_trends_report_q1_2021.pdf} (8 June, 2021)

\bibitem{report2}
Purplesec. "Cyber Security Statistics 2021." Available: \url{https://docs.apwg.org/reports/apwg_trends_report_q1_2021.pdf} (4 May, 2021)

\bibitem{report3}
APWG. "Phishing Activity Trends Report, 1st Quarter 2021." Available: \url{https://purplesec.us/resources/cyber-security-statistics/} (8 June, 2021)

\bibitem{report4}
Owasp. "Qrljacking." Available: \url{https://owasp.org/www-community/attacks/Qrljacking} (4 May, 2021)

\bibitem{report5}
Google. "Google 2 step verification." Available: \url{https://www.google.com/landing/2step/}

\bibitem{report6}
LastPass. "LastPass Authentication." Available: \url{https://www.lastpass.com/}

\bibitem{report7}
"Password Manager." Available: \url{https://www.airship.com/resources/explainer/push-notifications-explained/}

\bibitem{report8}
"Google Account Help." Available: \url{https://support.google.com/accounts/answer/7026266?co=GENIE.Platform%3DiOS&hl=en}

\bibitem{KSL19Q2}
Kaspersky. "Spam and Phishing in Q2 2019." Available: \url{https://securelist.com/spam-and-phishing-in-q2-2019/92379/} (Accessed on 21 October 2019)


\bibitem{report9}
  \emph{Biometric Advantages and Disadvantages}. Available: \url{https://www.sestek.com/2016/11/advantages-disadvantages-biometric-authentication/}

\bibitem{report10}
  \emph{Fingerprint Spoofing}. Available: \url{https://fortune.com/2016/02/24/fingerprint-spoofing-easy/}

\bibitem{report11}
  \emph{Face recognition facebook}. Available: \url{https://www.wired.com/2016/08/hackers-trick-facial-recognition-logins-photos-facebook-thanks-zuck/}

\bibitem{report12}
  \emph{Yubikey, U2F Fido standards}. Available: \url{https://www.yubico.com/authentication-standards/fido-u2f/}

\bibitem{report13}
  \emph{RSA SecurID}. Available: \url{https://www.webopedia.com/definitions/rsa-securid/}

\bibitem{report14}
  \emph{Yubikey failed}. Available: \url{https://www.yubico.com/blog/}

\bibitem{report15}
  \emph{MobileFaceNets: Efficient CNNs for Accurate RealTime Face Verification on Mobile Devices}. Available: \url{https://arxiv.org/ftp/arxiv/papers/1804/1804.07573.pdf}

\bibitem{report16}
  \emph{Access Control}. Available: \url{https://en.wikipedia.org/wiki/Access_control}

\bibitem{report17}
  \emph{Multi-factor Authentication}. Available: \url{https://searchsecurity.techtarget.com/definition/multifactor-authentication-MFA#:~:text=Multifactor%20authentication%20(MFA)%20is%20a,a%20login%20or%20other%20transaction.&text=Multifactor%20authentication%20is%20a%20core,identity%20and%20access%20management%20framework.}

\bibitem{report18}
  \emph{SAASPASS overview}. Available: \url{https://betanews.com/}

\bibitem{report19}
  \emph{SAASPASS 2FA scheme}. Available: \url{https://saaspass.com/technologies/proximity-instant-login-two-factor-authentication-beacon/}


\bibitem{report20}
  \emph{Tricipher, "Preventing man in the middle phishing attacks with multi-factor authentication"}. Available: \\
  \url{https://www.realwire.com/releases/tricipher-takes-identity-theft-} \\
  \url{prevention-mobile-with-affordable-portable-man-in-the-middle-protection}




\bibitem{report21}
  \emph{"2FA OTP-based scheme deprecated"}. Available: \url{https://searchsecurity.techtarget.com/definition/two-factor-authentication}

\bibitem{report22}
  \emph{"RSA SecurID compromised"}. Available: \url{https://arstechnica.com/information-technology/2011/06/rsa-finally-comes-clean-securid-is-compromised/}

\bibitem{report23}
  \emph{Avispa project}. Available: \url{http://www.avispa-project.org/}

\bibitem{report24}
  \emph{Deconstructing Alice and Bob}. Available: \url{http://www.avispa-project.org/papers/CVB-arspa05.pdf}

\bibitem{report25}
  \emph{Number of smartphone users worldwide from 2016 to 2026}. Available: \url{https://www.statista.com/statistics/330695/number-of-smartphone-users-worldwide/}

\bibitem{report26}
  \emph{Mobile Internet Usage Worldwide}. Available: \url{https://www.statista.com/topics/779/mobile-internet/#:~:text=In\%202020\%2C\%20the\%20number\%20of,mobile\%20device\%20to\%20go\%20online.}

\bibitem{report27}
  \emph{Fresh SmartPhone Statistics And What They Mean For You, NFC And The World}. Available: \url{https://nfc-forum.org/fresh-smartphone-statistics-and-what-they-mean-for-you-nfc-and-the-world/}

\bibitem{report28}
  \emph{4 Methods to Bypass two factor Authentication}. Available: \url{https://shahmeeramir.com/4-methods-to-bypass-two-factor-authentication-2b0075d9eb5f}

\bibitem{report29}
  Google. (2015). \emph{Stronger security for your google account}. Available: \url{https://www.google.com/landing/2step/}

\bibitem{report30}
  I. Barker. (2015). \emph{Saaspass makes two-factor authentication available to the masses}. Available: \url{https://betanews.com/2015/01/15/sa}

\bibitem{report31}
  M. M. P. Center. (2013). \emph{Browser extension hijacks Facebook profiles}. Available: \url{https://www.securityweek.com/malicious-firefox-chrome-extension-hijacks-facebook-profiles/}

\bibitem{report32}
  C. Hoffman. (2017). \emph{Beginner geek: Everything you need to know about browser extensions}. Available: \url{https://www.howtogeek.com/169080/beginner-geek-everything-you-need-to-know-about-browser-extensions/}

\bibitem{report33}
  \emph{Can iOS Safari access bluetooth device}. Available: \url{https://stackoverflow.com/questions/35072438/can-ios-safari-access-bluetooth-device}

\bibitem{report34}
  \emph{Security key for safer logins with touch in Facebook}. Available: \url{https://m.facebook.com/nt/screen/?params=\%7B\%22note_id\%22\%3A10157814544346886\%7D&path=\%2Fnotes\%2Fnote\%2F&_rdr}

\bibitem{report35}
  \emph{TRICIPHER, “Preventing man in the middle phishing attacks with multi-factor authentication," 2016}. Available: \\
  \url{https://www.helpnetsecurity.com/2005/03/22/tricipher-inc-announces-its-new-authentication-} \\
  \url{solution-protects-against-man-in-the-middle-phishing-attacks/}


\bibitem{report36}
FIDO Alliance, Available: \url{https://https://fidoalliance.org/specs/fido-v2.1-ps-20210615/fido-client-to-authenticator-protocol-v2.1-ps-20210615.html}

\bibitem{push1}
G. Varshney and Manoj Misra, "Push notification based login using BLE devices," in \textit{2017 2nd International conferences on Information Technology, Information Systems and Electrical Engineering (ICITISEE)}, 2017, pp. 479-484. DOI: 10.1109/ICITISEE.2017.8285554

\bibitem{graph1}
Ahmed Mahfouz, Tarek M. Mahmoud, and Ahmed Sharaf Eldin, "A survey on behavioral biometric authentication on smartphones," \textit{Journal of Information Security and Applications}, vol. 37, pp. 28-37, 2017. ISSN: 2214-2126. DOI: https://doi.org/10.1016/j.jisa.2017.10.002

\bibitem{phoneauth}
Alexei Czeskis, Michael Dietz, Tadayoshi Kohno, Dan Wallach, and Dirk Balfanz, "Strengthening User Authentication through Opportunistic Cryptographic Identity Assertions," in \textit{Proceedings of the 2012 ACM Conference on Computer and Communications Security} (CCS '12), New York, NY, USA, 2012, pp. 404–414. ISBN: 9781450316514. DOI: 10.1145/2382196.2382240

\bibitem{mukho}
Zhenfeng Xu, Huajun Yin, Pei Xiong, Chuan Wan, and Qing Liu,
\emph{Short-term responses of Picea asperata seedlings of different ages grown in two contrasting forest ecosystems to experimental warming},
\emph{Environmental and Experimental Botany},
vol. 77, pp. 1-11, 2012.
\url{https://www.sciencedirect.com/science/article/pii/S0098847211002620}

\bibitem{dodson}
Zhenfeng Xu, Huajun Yin, Pei Xiong, Chuan Wan, and Qing Liu,
\emph{Short-term responses of Picea asperata seedlings of different ages grown in two contrasting forest ecosystems to experimental warming},
\emph{Environmental and Experimental Botany},
vol. 77, pp. 1-11, 2012.
\url{https://www.sciencedirect.com/science/article/pii/S0098847211002620}

\bibitem{zhu}
Bin B. Zhu, Jeff Yan, Guanbo Bao, Maowei Yang, and N. Xu,
\emph{Captcha as Graphical Passwords—A New Security Primitive Based on Hard AI Problems},
\emph{IEEE Transactions on Information Forensics and Security},
vol. 9, pp. 891-904, 2014.

\bibitem{camauth}
Mengjun Xie, Yanyan Li, Kenji Yoshigoe, Remzi Seker, and Jiang Bian,
\emph{CamAuth: Securing Web Authentication with Camera},
\emph{2015 IEEE 16th International Symposium on High Assurance Systems Engineering},
pp. 232-239, 2015.
DOI: 10.1109/HASE.2015.41

\bibitem{spoof}
Luka Malisa, Kari Kostiainen, and Srdjan Capkun,
\emph{Detecting Mobile Application Spoofing Attacks by Leveraging User Visual Similarity Perception},
\emph{Proceedings of the Seventh ACM on Conference on Data and Application Security and Privacy},
pp. 289-300, 2017.
DOI: 10.1145/3029806.3029819
\bibitem{saxena}
Nitesh Saxena, Jan-Erik Ekberg, Kari Kostiainen, and N. Asokan,
\emph{Secure Device Pairing Based on a Visual Channel: Design and Usability Study},
\emph{IEEE Transactions on Information Forensics and Security},
vol. 6, pp. 28-38, April 2011.
DOI: 10.1109/TIFS.2010.2096217

\bibitem{ali}
Inayat Ali, Sonia Sabir, and Zahid Ullah,
\emph{Internet of Things Security, Device Authentication and Access Control: A Review},
arXiv preprint arXiv:1901.07309, 2019.
\url{https://arxiv.org/abs/1901.07309}

\bibitem{camtalk}
Mengjun Xie, Liang Hao, K. Yoshigoe, and Jiang Bian,
\emph{CamTalk: A Bidirectional Light Communications Framework for Secure Communications on Smartphones},
\emph{Proceedings of the 2013 IEEE 16th International Symposium on High Assurance Systems Engineering},
vol. 127, pp. 35-52, September 2013.
DOI: 10.1007/978-3-319-04283-1\_3


\bibitem{phool}
Bryan Parno, Cynthia Kuo, and Adrian Perrig,
\emph{Phoolproof Phishing Prevention},
\emph{Financial Cryptography and Data Security},
pp. 1-19, Springer Berlin Heidelberg, 2006.
ISBN: 978-3-540-46256-9

\bibitem{study}
G. Varshney, Manoj Misra, and Pradeep Atrey,
\emph{Secure authentication scheme to thwart RT MITM, CR MITM and malicious browser extension based phishing attacks},
\emph{Journal of Information Security and Applications},
vol. 42, pp. 1-17, 2018.
DOI: 10.1016/j.jisa.2018.07.001
\url{https://www.sciencedirect.com/science/article/pii/S2214212618300140}

\bibitem{c1}
Chun-Ying Huang, Shang-Pin Ma, and Kuan-Ta Chen,
\emph{Using one-time passwords to prevent password phishing attacks},
\emph{Journal of Network and Computer Applications},
vol. 34, no. 4, pp. 1292-1301, 2011.
\emph{Advanced Topics in Cloud Computing}.
DOI: 10.1016/j.jnca.2011.02.004
\url{https://www.sciencedirect.com/science/article/pii/S1084804511000427}


\bibitem{c2}
Vasileios Mavroeidis and Mathew Nicho,
\emph{Quick Response Code Secure: A Cryptographically Secure Anti-Phishing Tool for QR Code Attacks},
\emph{Computer Network Security},
Springer International Publishing, Cham, 2017, pp. 313--324.
ISBN: 978-3-319-65127-9

\bibitem{c3}
Inayat Ali, Sonia Sabir, and Zahid Ullah,
\emph{Product Authentication Using QR Codes: A Mobile Application to Combat Counterfeiting},
Available: \url{https://doi.org/10.1007/s11277-016-3374-x}

\bibitem{c4}
\emph{Securing SMS Based One Time Password Technique from Man in the Middle Attack},
Available: \url{https://arXiv:1405.4828}

\bibitem{c5}
Jingai Xu, Jiang Qi, and Ye Xi,
\emph{OTP bidirectional authentication scheme based on MAC address},
\emph{2016 2nd IEEE International Conference on Computer and Communications (ICCC)}, 2016, pp. 1148--1152.
DOI: 10.1109/CompComm.2016.7924884

\bibitem{c6}
K. Sudhakar, S. Srikanth, and M. Sethuraman,
\emph{Secured mutual authentication between two entities},
\emph{2015 IEEE 9th International Conference on Intelligent Systems and Control (ISCO)}, 2015, pp. 1--5.
DOI: 10.1109/ISCO.2015.7282338

\bibitem{c7}
Yunlim Ku et al.,
\emph{Two-factor authentication system based on extended OTP mechanism},
\emph{International Journal of Computer Mathematics},
vol. 90, no. 12, pp. 2515--2529, 2013.
Publisher: Taylor \& Francis
DOI: 10.1080/00207160.2012.748901
URL: \url{https://doi.org/10.1080/00207160.2012.748901}


\bibitem{c8}
\emph{The implementation of two-factor web authentication system based on facial recognition},
Available: \url{https://doi.org/10.18844/gjcs.v7i2.3448}

\bibitem{c9}
Baber Aslam, Lei Wu, and Cliff C. Zou,
\emph{PwdIP-Hash: A Lightweight Solution to Phishing and Pharming Attacks},
\emph{2010 Ninth IEEE International Symposium on Network Computing and Applications}, 2010, pp. 198--203.
DOI: 10.1109/NCA.2010.35

\bibitem{c10}
G. Varshney, Manoj Misra, and Pradeep Atrey,
\emph{A new secure authentication scheme for web login using BLE smart devices},
\emph{2017 11th IEEE International Conference on Anti-counterfeiting, Security, and Identification (ASID)}, 2017, pp. 95--98.
DOI: 10.1109/ICASID.2017.8285751

\bibitem{Deng_2022}
Jiankang Deng, Jia Guo, Jing Yang, Niannan Xue, Irene Kotsia, Stefanos Zafeiriou.
\newblock "ArcFace: Additive Angular Margin Loss for Deep Face Recognition."
\newblock \textit{IEEE Transactions on Pattern Analysis and Machine Intelligence}, vol. 44, no. 10, pp. 5962–5979, Oct. 2022.
\newblock ISSN: 1939-3539.
\newblock DOI: \url{10.1109/TPAMI.2021.3087709}.
\newblock Available at: \url{http://dx.doi.org/10.1109/TPAMI.2021.3087709}.



\bibitem{wiki_ctap}
Wikipedia contributors. "Client to Authenticator Protocol." Wikipedia, The Free Encyclopedia. Available online: \url{https://en.wikipedia.org/wiki/Client_to_Authenticator_Protocol}. 




\end{thebibliography}
\end{document}